\documentstyle[aaspp4, psfig]{article}

\begin{document}

\title{Orbital Evolution of Planets\\
Embedded in a Planetesimal Disk\vspace*{0.8in}}

\author{Joseph M. Hahn}
\affil{Lunar and Planetary Institute, 3600 Bay Area Boulevard,
Houston, TX 77058--1113\\
email: hahn@lpi.jsc.nasa.gov\\
phone: 281--486--2113\\
fax: 281--486--2162}

\vspace*{0.5in}

\author{Renu Malhotra}
\affil{Lunar and Planetary Institute, 3600 Bay Area Boulevard,
Houston, TX 77058--1113\\
email: renu@lpi.jsc.nasa.gov\\
phone: 281--486--2114\\
fax: 281--486--2162}

\vspace*{0.5in}

\begin{center}
Accepted for publication in {\it Astronomical Journal}.\\
February 16, 1999.\\
LPI Contribution No.\ 966.
\vspace{0.5in}

\end{center}

\newpage
\begin{abstract}
\baselineskip=15pt

The existence of the Oort Comet Cloud, the Kuiper Belt, and plausible
inefficiencies in planetary core formation, all
suggest that there was once a residual planetesimal disk of mass
$\sim10$--100 M$_\oplus$ in the vicinity of the giant planets
following their formation. Since removal of this disk
requires an exchange of orbital energy and angular momentum
with the planets, significant planetary migration can ensue.
The planet migration phenomenon is examined
numerically by evolving the orbits of the giant planets while they
are embedded in a planetesimal disk having a mass of
$M_D=10$ to 200 M$_\oplus$. We find that Saturn, Uranus, and Neptune
evolve radially outwards as they scatter the planetesimals, while
Jupiter's orbit shrinks as it ejects mass. Higher--mass disks result in
more rapid and extensive planet migration. If orbit expansion and
resonance trapping by Neptune is invoked to explain the eccentricities
of Pluto and its cohort of Kuiper Belt Objects at Neptune's 3:2
mean--motion resonance, then our simulations suggest that a disk mass
of order $M_D\sim50$ M$_\oplus$ is required to expand Neptune's orbit
by $\Delta a\sim7$ AU in order to
pump up Plutino eccentricities to $e\sim0.3$.
Such planet migration implies that the initial Solar System
was more compact in the past, with the Jupiter--Neptune separation
having been smaller by about $30\%$.

We discuss the fate of the remnants of the primordial planetesimal disk.
We point out that most of the planetesimal disk beyond
Neptune's 2:1 resonance should reside in nearly circular,
low--inclination orbits, unless there are (or were) additional unseen,
distant perturbers.
The planetesimal disk is also the source of the Oort Cloud of comets.
Using the results of our simulations together with a simple treatment
of Oort Cloud dynamics, we estimate that $\sim12$ M$_\oplus$ of disk
material was initially deposited in the Oort Cloud,
of which $\sim4$ M$_\oplus$ will persist over the age of the Solar System.
The majority of these comets originated from the Saturn--Neptune
region of the solar nebula. 

\end{abstract}


\newpage
\section{Introduction}
\label{intro}

It is generally accepted that a final distinct stage in the formation
of our planetary system consisted of 
the clearing of a residual planetesimal population by the
gravitational perturbations of fully--formed giant planets.
The formation of the Oort Cloud, which is
a spherical swarm of comets orbiting the Sun at
distances of $\sim10^{3\ \mbox{\scriptsize to\ } 5}$ AU,
is thought to be a product of this stage of Solar System formation.
Mass estimates for the Oort Cloud are in the range
$\sim10$ to $100 M_\oplus$ (\cite{W96}).
Another remnant of the planetesimal disk is the recently discovered
population of the Kuiper Belt Objects (KBOs) that orbit beyond Neptune.
Although the mass of the observable portion of the Kuiper Belt
is only $\sim0.26$ M$_\oplus$ (\cite{Jetal98}),
planetesimal accretion models require an initial mass that is of order
$\sim35$ M$_\oplus$ for the assembly of Pluto and the
$R\sim100$ km--sized KBOs in the 30--50 AU zone (\cite{SC97},
\cite{KL98}). Noting also that the solid cores of the giant
planets are of order $\sim10$ M$_\oplus$, and that core formation is
not likely to have been $100\%$ efficient, it is quite plausible that
there was still a residual planetesimal disk of mass
$\sim10$---100 M$_\oplus$ in the vicinity of
the giant planets after they formed.
The eventual removal of this mass via gravitational scattering
by the giant planets could have caused significant evolution of the
planetary orbits, such that the presently observed orbital configuration
of the Jovian planets is considerably altered from that which
was obtained soon after their formation.

Not only is an early epoch of planet migration plausible based upon
formation considerations, but
recent advances in our knowledge of the outer Solar System provide
new motivation for studying this process.
The determination of reasonably reliable orbits for several dozen KBOs
has revealed that their distribution is quite non--uniform:
there is a near--complete dearth of KBOs having semimajor axes $a$
interior to Neptune's 3:2 mean--motion resonance at 39.4 AU, and
there is a prominent concentration of objects at the 3:2 resonance with
moderately large eccentricities $e\sim0.1$---0.35 (\cite{Jetal98}).
Several other KBOs orbit at the 4:3, 5:3, and at the 2:1 resonances,
all with moderately large eccentricities.
There is also a non--resonant KBO population beyond the 3:2 resonance
which extends out to 47 AU; these objects exist in more circular orbits
with $e\lesssim0.1$.
This peculiar orbital distribution supports the hypothesis
that Neptune's orbit migrated radially outwards, sweeping
the primordial Kuiper Belt with that planet's mean--motion resonances,
and capturing Pluto as well as a cohort of KBOs at those resonances
(\cite{RM93}, \cite{RM95}).
Both the semimajor axes and the eccentricities of captured bodies
would have grown concurrently with the planet's orbital expansion.
The inclinations of KBOs can also become excited when a vertical
secular resonance sweeps past (\cite{RM98}).
Analysis of the resonance--sweeping mechanism shows that
if planet migration were responsible for the eccentric and inclined
orbits of Pluto and the other KBOs at the 3:2 resonance, then
Neptune's orbit must have expanded by $5\lesssim\Delta a\lesssim10$ AU
on a timescale of order $\tau_m\sim10^7$ years or longer (\cite{RM93},
\cite{RM98}).  \cite{RM97} estimated that the gravitational clearing
of a residual planetesimal disk having a mass of about 35 M$_\oplus$
distributed in the vicinity of Uranus and Neptune
would expand Neptune's orbit by $\Delta a\sim5$ AU.
The planet--migration/resonance--sweeping hypothesis
not only accounts for the abundance of eccentric, inclined KBOs locked
in orbital resonance with Neptune, but it also accounts for the
lack of low--eccentricity orbits in the 3:2, the depleted
region interior to the 3:2 as well as the excited $e\sim0.1$ state of
the non--resonant KBOs. It is thus possible that the
characteristics of the orbital migration history
of Neptune (and, by extension, the other giant planets) is preserved
in the details of the KBO orbital distribution.

Orbital migration was first noted in the giant planet accretion
models of \cite{FI(84)} which showed that the orbits of a
growing proto-- Uranus and Neptune could shift by $\sim5$--10 AU while
embedded in a planetesimal disk having a mass of order
$\sim100$ M$_\oplus$
(\cite{FI84}, \cite{FI96}). \cite{FI(84)} argue that Uranus and Neptune
would preferentially gain angular momentum (and thus expand their
orbits) by scattering planetesimals and lowering
their perihelia down to Jupiter and Saturn. Jupiter, being an effective
planetesimal ejector, would shrink its orbit due to the resulting
energy and angular momentum losses. However it should be noted that
this argument, while plausible, is not fully supported
by their numerical simulations. Those simulations show net orbital
expansion by Uranus and Neptune in some instances and orbital decay in
others. Individual runs show considerable to--and--fro motion by
Neptune. Such evolution would only stir up the planetesimals
rather than trap them at resonances.
We note that these simulations were not fully self--consistent since
gravitational interactions were considered only
between planets and planetesimals that were on crossing orbits;
long range forces between the planets and the planetesimals were neglected,
as were the mutual planet-planet perturbations.
Also, the giant planets' gravitational
cross--sections were artificially increased in order to speed up the
system's evolution.  This approximate treatment of planet-planetesimal
dynamics was necessitated partly due to the limited computer resources
available at the time. A more realistic treatment of
disk clearing and planet--migration is clearly warranted.

The present work revisits the planet--migration scenario using
more sophisticated numerical techniques to
evolve a system of giant planets that is embedded
in a disk composed of many low--mass particles.
In Section \ref{method}, we describe the model and test results in detail.
Primary results on the giant planets' orbital evolution
with disks having different initial masses are given in Section
\ref{migration}.
Since the planetesimal disk is the source of Oort Cloud,
we discuss the implications of our results for the formation
of the Oort Cloud in Section \ref{oort_cloud}.
Particles that manage to avoid ejection will
reside in what are sometimes called the classical and the
scattered disks, which are characterized in
Section \ref{scat_disk}.
The role of unmodeled effects such as
disk self--gravity and the role of spiral density waves
are discussed in Section \ref{waves},
and conclusions are given in Section \ref{conclusions}.

\section{Numerical Method}
\label{method}

Our model consists of the Sun, the four giant planets Jupiter, Saturn,
Uranus, Neptune, and a population of numerous $(N)$ low--mass
particles initially distributed in a disk. Ideally, one would like to
have $N\sim10^{10}$ to follow the evolution of this system
with a completely self-consistent calculation of the mutual
gravitational perturbations and also include the external forcing due
to the galactic tide. However such a calculation exceeds current
computational resources. Instead, as we describe below,
we use a fast orbit integrator together with several
simplifications to obtain an approximation to the ideal
within a reasonable amount of computing time.

In our simulations, all bodies are treated as point masses,
thus the possibility of collisions or accretion is neglected.
This approximation is justified during the late
stages of giant planet formation when the frequency of planet-particle
collisions is smaller than the scattering frequency by a factor of the
planet's physical radius/Hill radius squared, which ranges from 
$10^{-6}$ for Jupiter to $10^{-8}$ for Neptune.

The model includes the mutual gravitational forces exerted between
the planets as well as the forces between the planets
and the low--mass particles, but the forces amongst the particles
themselves are neglected. This approximation is employed to reduce
the computational expense, and it also eliminates
the unphysical self--stirring that would otherwise occur in the
model disks considered below that have masses of $M_D=10$ to 200
M$_\oplus$ distributed among just $N=1000$ particles.
Had the particle-particle interactions been included, the relatively
small number of $N$ massive disk particles would quickly stir
themselves up so much that the resulting system would hardly resemble
a planet--forming particle disk. Although this difficulty is avoided by
neglecting the particle--particle forces, this approximation also
precludes the possibility that the planets might generate spiral
density waves in the disk (see Section \ref{waves}).

Dynamical models have shown that particles
scattered into wide orbits with semimajor axes $a\gtrsim3000$ AU
become decoupled from the planets due to the galactic tide
(\cite{DQT87}; see also Section \ref{oort_cloud}). These distant
bodies are deposited in the Oort Cloud, which is effected
here by simply removing particles from the system when $a>3000$ AU.

A second-order mixed variable symplectic (MVS) 
mapping is used to rapidly advance
the heliocentric positions and velocities of the planets and
low--mass particles as they interact in the Sun's gravitational field.
Our implementation follows the algorithm of \cite{WH(91)} with the
improvements of \cite{ST(92)}. A fixed step size
of $\Delta t=0.4$ years is used, which is sufficiently short to
resolve the orbits of particles that evolve down to perihelia of
$q\simeq3.5$ AU as well as most encounters between particles and
planets. For the latter, the dynamical timescale of a typical
scattering event is\footnote{This encounter time is defined as the
time to change the particle's true anomaly from $-\pi/2$ to $\pi/2$
for a planet--centered parabolic orbit having a pericenter distance
$q$ from the planet.}
\begin{equation}
T^\star\simeq\frac{8}{3}\sqrt{\frac{2q^3}{GM_p}}
\end{equation}
where $M_p$ is the scattering planet's mass and $q$ is the
particle's closest approach distance. Evaluating $T^\star$
with $q$ set to each planet's Hill sphere radius shows that
$T^\star\simeq4.1$ years at Jupiter, 10 years at Saturn, 29 years at
Uranus, and 57 years at Neptune. Thus the integration timestep of
$\Delta t=0.4$ years is at worst 1/10 of the dynamical time for
orbits that graze Jupiter's Hill sphere. However an MVS algorithm
that employs a fixed step size will fail to resolve very close
planet--particle encounters that penetrate well within a planet's Hill
sphere, and will also fail to correctly evolve
a very eccentric orbit when the perihelion passage
is not well time--sampled (\cite{RH98}). These difficulties are
mitigated by the using the following approximations.

({\it i.}) During a close encounter between a particle and a planet,
two--body trajectories are adopted for their relative
motion whenever a particle passes sufficiently close to a planet.
Experimentation shows that a ``sufficiently close''
encounter is one that changes
the particle's fractional distance from the planet
by more than 50\%, or varies its planet--centered angular coordinate
by more than 90$^\circ$, during the time step $\Delta t$. For particles
that are initially in low--eccentricity heliocentric orbits,
the two--body approximation is triggered only when a particle 
approaches within
about $10\%$ of the planet's Hill sphere. For
more distant encounters, particle trajectories are in fact evolved
with greater accuracy using the standard MVS mapping.

({\it ii.}) When a particle evolves into an orbit
of sufficiently high eccentricity,
its motion during perihelion passage may be poorly
time-sampled and the MVS mapping can produce unphysical evolution.
For the step size employed here, significant errors
accrue in orbits having perihelia
$q\lesssim3.5$ AU\footnote{It should be noted that the symplectic
integrators of \cite{LD(94)} and \cite{DLL(98)} do not suffer this
instability when {\it massless} test particles achieve low perihelia
orbits (H.\ Levison, personal communication). However the intervention
described above is still required since there is no fully
symplectic algorithm that can compute low--perihelia trajectories
when the particles carry mass.}.
Proper treatment of such orbits is not of minor import since 
approximately 30\% of the disk particles cycle through the
inner Solar System with heliocentric distances of $r<3.5$ AU. 
A simple, but impractical solution, is to
use a step size small enough to resolve all perihelion passages.
However a more practical approach consists of turning off all
planetary perturbations while a particle travels
interior to a heliocentric distance of $r<3.5$ AU; this results in
a piecewise Keplerian trajectory that is approximately correct.
With this approximation, the particle's perturbations upon the
planets are still not fully time--sampled, which causes
slow drifts in the system's total energy and angular momentum.
Numerical experimentation has shown that the most economical
and robust solution to this problem requires turning off the
particle's perturbations upon the planets
while its {\sl perihelion} distance is $q<3.5$ AU.

The remainder of this section describes a few critical tests of the
algorithm in order to demonstrate that the approximations made here
do not introduce any artificial orbital evolution of the planets and
the particles that is of any significance.

The quality of the orbit integrations may be illustrated
with the restricted three--body problem that consists of
a massless particle perturbed by a planet on a circular orbit
about the Sun. For this system the particle's Jacobi integral
$J=E-L_z\Omega_p$ is conserved, where $E$ is the particle's energy,
$L_z$ is its component of angular momentum perpendicular to the
planet's orbit, and $\Omega_p$ is the planet's mean motion.
Figure \ref{delta J} shows the mean fractional variations $\Delta J/J$
that result when integrating 100 particles for $10^4$ years
with a Jupiter-mass planet at $a_p=5$ AU.
Squares are shown for those particles that start in rather
circular orbits near and beyond the planet; these experience a
$\Delta J/J\lesssim10^{-6}$ that decreases with distance
provided they have not already made a close approach to Jupiter. Filled
squares indicate particles that
approach closer than a Hill distance to Jupiter; these
experience larger variations in $J$ due to the close--encounter
approximation used here. The remaining particles (crosses and circles)
are all on Jupiter--crossing orbits with their perihelia ranging
over $0.05<q<6$ AU. Crosses
are shown for particles that do not encounter Jupiter,
whereas filled circles represent particles that do encounter
the planet. The crosses show that $\Delta J/J$ grows inwards of
Jupiter's orbit due to the poor time--sampling of the particles'
perihelion passages. However the error $\Delta J/J$ levels
off where the integrator turns off the planet's perturbation
and the particle motions are temporarily Keplerian.
Failure to implement this procedure would have instead
yielded disastrous results inwards of about
$q\simeq1$ AU. Shrinking the step size $\Delta t$ would of course
reduce the error in $J$ but at the expense of precious
computer cycles.

An alternate scheme that avoids the growth of numerical errors among
low--$q$ particles is to simply remove them when they drop down to a
heliocentric distance of $r<3.5$ AU. Although this procedure inhibits
error growth, it is less than desirable since it introduces an
unphysical (and very hungry) mass sink to the system. When employing
this procedure to the simulations of planet--migration
described in Section \ref{migration} (which employ the Keplerian
approximation whenever $r<3.5$ AU), we find that the orbital
migration of Saturn, Uranus, and Neptune proceeds radially outwards at
similar rates. However the removal of $r<3.5$ AU particles
results in an artificial sunward mass--flux
that causes Jupiter's orbit to expand rather than shrink.
Nonetheless, the fact that the orbital evolution of the outer
giant--planets' is rather insensitive to the treatment of
low--$q$ particles adds confidence that the Keplerian approximation
employed here is not driving the planet--migration reported in Section
\ref{migration}. However enforcing the particles'
Keplerian motion inside of $r<3.5$ AU is still needed in order to
accurately model Jupiter's orbital history.

In the absence of any close encounters between particles and
planets, the MVS integrator preserves the system's
integrals in the expected manner. However Fig. \ref{delta J} shows
that the close encounter algorithm used here
will not preserve a particle's $J$ to better than 1 part in $10^3$
should it pass nearer than a Hill distance of a planet. This
numerical error
results in an unphysical diffusion of particle trajectories.
This is a concern since the planet migration
phenomenon is sensitive to the flux of particles that
are scattered and exchanged amongst the planets
(\cite{FI84}, \cite{RM93}).

To judge the effects of numerical diffusion, the restricted
three--body problem is again integrated for a system of
100 massless particles that orbit in the vicinity of a Neptune--mass
planet in a circular orbit at $a_p=30$ AU.
The particles' initial semi-major axes range over
$28.7<a<31.5$ AU and have small initial eccentricities $e$ and 
inclinations $i$. For this system, each particle's dimensionless
Jacobi integral,
\begin{equation}
\label{jacobi}
J=\frac{a_p}{a}+2\sqrt{\frac{a}{a_p}(1-e^2)}\cos i+{\cal O}(\mu),
\end{equation}
is a conserved quantity. Here, a particle's semimajor
axis $a$ is in units of $a_p$ and $\mu=5.15\times10^{-5}$ is
Neptune's mass in solar units.
In this test run, the particles initially have a mean Jacobi
integral $\langle J\rangle\simeq2.998$ with a dispersion
$\sigma_J=5\times10^{-4}$.
To obtain a physically meaningful measure of the
numerical diffusion that occurs in the model, define a threshold value 
$J_U=2.989$ which corresponds to an orbit having a
perihelion at 20 AU (near Uranus' orbit) and an aphelion at
30 AU (near Neptune's orbit) so that $a/a_p=5/6$ and $e=1/5$.
Although this orbit is dynamically forbidden to
the test particles, numerical diffusion might allow particles to cross
Uranus' orbit which would result in an unphysical exchange of particles
between Neptune and Uranus. The system is integrated for
$5\times10^7$ years, and at the end of the run 11 particles have
diffused into forbidden Uranus-crossing orbits that have $J<J_U$.
The timescale to diffuse into crossing orbits, $t_d$, is obtained
from Fig. \ref{J diffusion} which shows the particles' average $J$
and their dispersion $\sigma_J$ versus time. The characteristic
diffusion timescale for particles to diffuse across the Jacobi `gap',
{\it i.e.}, the time when $|J-\sigma_J|<J_U$, is
$t_d\simeq2.5\times10^7$ years.

However it should be noted that a particle's $J$ is not
conserved in a multi--planet system.
For example, adding a Jupiter at 5 AU to the above simulation
will drive these particles into Uranus-crossing
orbits having $J<J_U$ at a rate that is about 10 times faster.
Since the numerical diffusion rate is considerably slower than the
dynamical diffusion that occurs in a multi--planet system,
we conclude that the transport of particles between the planets
due to numerical diffusion is not significant in these simulations.

As already noted, 
it is the close encounters between the planets and the 
planetesimals, and their concomitant exchange of angular momentum,
which drive the planet migration process.
In our simulations, the planet-particle relative motion during
very close  encounters is not computed exactly, so any small
error in the angular momentum exchange is also reflected in the
recoiling planet's orbit. The final test discussed below
verifies that the errors in this angular momentum
exchange are in fact too small to drive the planet--migration
described in Section \ref{migration}.

The motion of Neptune as well as several hundred
particles of infinitesimal mass $m$ are integrated at
the usual step size $\Delta t=0.4$ years for 200 years.
The same experiment
is then repeated using a step size 50 times smaller. Since
integration errors decrease with step size,
this latter run may be regarded as a much more exact representation
of the particle trajectories.
Upon differencing the two runs, the error in each particle's
velocity relative to the planet, $\delta {\bf V}$,
is calculated as the particle exits the planet's Hill sphere
(see Fig.\ \ref{dV}a). The Figure shows that most encounters occur
at the periphery of the planet's Hill sphere which result in rather
small relative velocity errors. However the very close
encounters having periapse $q\lesssim0.1R_H$,
which account for $6\%$ of these encounters, result in sizable
velocity errors that are of order $\delta V/V\sim0.1$.
The error in the specific angular momentum $\ell_p$ exchanged between
the planet and a scattered particle is
$\delta \ell_p=-m{\bf r}_p\times\delta{\bf V}/(M_p+m)$,
where ${\bf r}_p$ is the planet's heliocentric coordinate and $M_p$ is
its mass. Figure \ref{dV}b shows a histogram of the $z$--component of
the errors $\delta \ell_p$ which are distributed about zero.
The net specific angular momentum exchanged
between the planet and the particle swarm is the sum
$L=|\sum\ell_p|$, and its root--mean--square error is
$\delta L=\sqrt{\sum\delta\ell_p^2}$. For the encounters shown in
Fig.\ \ref{dV}, the fractional error in the angular
momentum exchanged is only $\delta L/L=0.016$. The above procedure is
also repeated for encounters at Jupiter, which are less well
time--resolved, and yields a larger fractional error
$\delta L/L=0.12$. It should be noted that these
fractional errors decrease for higher relative velocities, which
reflects the fact that the two--body approximation, which neglects
the Sun's gravity, becomes more accurate at faster encounters.
Thus the error in the total angular momentum exchanged between the
planet and neighboring particles will steadily decrease as the
planet heats up the particle disk. Since these fractional errors are
small, we conclude that the close encounter
approximation used here does not drive the planet migration
described below.

\section{Simulations of Planet Migration}
\label{migration}

\subsection{Initial conditions}

In the simulations reported here, we adopted initial planet orbits
similar to those used in previous investigations
of planet migration, in which the semimajor axes of Jupiter, Saturn,
Uranus, and Neptune are displaced from their present orbits by
respective amounts $\Delta a=+0.2$, -0.8, -3.0, and -7.0 AU
(\cite{RM95}). The planets are assumed to have their present masses.
The initial planetesimal disk is composed of 1000 equal--mass
particles distributed in orbits of $10<a<50$ AU such that the disk's
inner edge lies just exterior to Saturn and the outer
edge lies just beyond the present location of Neptune's 2:1 resonance.
The disk surface density $\sigma$ varies as $a^{-1}$, 
$\sigma(a)=4.0\times10^{-3}(M_D/\mbox{AU}^2)(\mbox{1 AU}/a)$,
where $M_D$ is the total disk mass.  Thus, approximately
two--thirds of the disk starts
exterior to Neptune's initial orbit.
Four separate simulations are presented below in which the disk's
initial mass is $M_D=10$, 50, 100, and 200 M$_\oplus$,
and the individual disk particles have masses
$m=0.01$, 0.05, 0.1 and 0.2 M$_\oplus$. An additional 50
massless particles are also distributed between $50<a<100$ AU in
order to assess the degree of perturbation of a hypothetical
part of the Kuiper Belt extending well past that
which is presently observable. 

It should be noted that accretion models advocate
an initial disk containing several tens of Earth--masses
in the 30--50 AU zone in order to form Pluto and QB$_1$--type
Kuiper Belt Objects (\cite{SC97}, \cite{KL98}), a scenario
that is bracketed by the models explored here.

For those particles that initially lie far from any planets, initial
eccentricities of $e_d=0.01$ and inclinations $i_d=e_d/2$ are adopted.
However particles having a semimajor axis near a planet will already
have experienced a history of stirring that results in particle
dispersion velocities $v_d$ of the order of the planet's escape
velocity at its Hill sphere radius $R_H$ (\cite{IM93}),
where
\begin{equation}
v_d\sim\sqrt{2GM/R_H} \quad
\mbox{and} \quad
R_H=(M/3M_\odot)^{1/3}a,
\end{equation}
$G$ is the gravitation constant, $M$ the planet's mass,
and $M_\odot$ the solar mass. Assuming the particles'
inclinations are half their eccentricities, the particles'
dispersion velocity is $v_d\simeq\sqrt{(e_d^2+i_d^2)GM_\odot/a}\sim
e_d\sqrt{5GM_\odot/4a}$ which
corresponds to an eccentricity of $e_d\sim\sqrt{24/5}(R_H/a)$.
With $R_H/a\simeq0.025$ for both Uranus and Neptune,
the particles starting in each planet's heated zone
have initial $e_d=0.06$ and $i_d=1.7^\circ$.
The adopted half-width of each planet's heated zone is
simply each planet's `feeding zone' of $\Delta a=2\sqrt{3}R_H$
(e.g.\ \cite{IM93}) such that $\Delta a=1.4$ AU for Uranus and 2.0 AU
for Neptune when evaluated at their initial heliocentric distances.

For a planet that is embedded in a swarm of identical
particles of mass $m$, dynamical friction will tend to seek
an equipartition of energies in the system's epicyclic motions. Again 
assuming a planet's inclination obeys $i_p=e_p/2$,
its initial eccentricity is then $e_p=e_d\sqrt{m/M}$,
where $e_d$ are the eccentricities of particles in the heated zone.
Since Jupiter and Saturn start interior to the particle disk,
initially circular and coplanar orbits are adopted for these two
planets.

\subsection{Results}

Figure \ref{a(t)} shows the orbital histories of
the four giant planets as they scatter the surrounding disk particles
in each of the four simulations.
We find that the lowest--mass disk, $M_D=10M_\oplus$,
yields little evolution in the planets' orbits, but the higher--mass
disks result in significant radial displacements of the planets
during the 30 Myr runs.  These simulations confirm the expectation
that Saturn, Uranus and Neptune
migrate radially outwards while Jupiter migrates slightly
inwards. For a given disk mass, the magnitude of radial
migration is largest for Neptune and successively less so for the
interior planets. As might be expected,
planetesimal disks of greater mass result in
planet migration that is larger in magnitude and more rapid,
and also more stochastic owing to the individual particles'
greater mass. Although the planets' orbital eccentricities and
inclinations remain small, there is a clear trend towards larger $e$'s
and $i$'s at larger disk masses. Since their final $e$'s and
$i$'s are largely determined by the numbers and masses of disk
particles used in the simulations, their final state have little
relation to the giant planets current $e$ and $i$ configurations.
It is worth noting that, in the high--mass disk simulations,
the planets pass through a few mutual low--order
mean--motion resonances, but the planets' orbits do not
persist in any resonance-locked configurations. 
This is not entirely surprising as the general trend in
the orbital evolution is such that the planets are driven
towards greater mutual orbital separation, 
which is not conducive to maintaining a resonance libration
(\cite{DMM88}).

Figure \ref{ea} shows the state of the $M_D=50$ M$_\oplus$ system
at logarithmic time intervals. Large dots
indicate the planets' eccentricities and semimajor axes while
small dots/crosses denote particles that have/have not had a close
approach to a planet. Those particles scattered by the planets yet
still bound to the Sun tend to have perihelia between the orbits of
Saturn and Neptune, as indicated by the two curves.
It should be noted that in all of the simulations reported here, the
orbital migration of the planets has not ceased by the end of the runs,
and that further planet--migration will continue (albeit more slowly)
past 30 Myr. The simulation of the $M_D=50$ M$_\oplus$ disk is
extended to 50 Myr in Fig.\ \ref{a(t)_50}, which shows Neptune
slowly expanding its orbit out to 30 AU. These results suggest
that in order to actually `park' Neptune at $a=30$ AU requires an
adjustment of parameters towards a slightly lower disk mass, and/or a
steeper gradient in the disk's surface density profile, and/or 
an outer disk edge that lies closer than 50 AU.

\subsection{Discussion}
\label{back_torque}

The disk particles used in all of these simulations are sufficiently
massive that their perturbations upon Neptune result in non--adiabatic
expansion of that planet's orbit. However it is evident in Fig.\
\ref{a(t)} that planet migration is smoother when the disk particles
have lower mass. Thus a more realistic simulation employing larger
numbers of particles will better resolve the disk--planet
perturbations. We expect that such simulations will exhibit
orbit migration which proceeds more smoothly. Nearly adiabatic orbital
migration is in fact required if Neptune is to efficiently capture
particles at its exterior mean--motion resonances.
Since smooth outward expansion was not realized in these simulations,
Neptune did not capture any particles at its mean--motion resonances.
Thus the simulated disk's end state cannot be directly
compared to the delicate resonant structure that is observed in the
Kuiper Belt. Nevertheless, these simulations do provide useful
constraints on the likely mass of the initial debris disk that
may have been present during an early epoch of planet migration.

Figure \ref{da} reports the radial displacement $\Delta a$
experienced by each planet after $t=3\times10^7$ years
as a function of the initial disk mass $M_D$. As noted above,
Jupiter migrates sunwards while the other
planets migrate outwards in each simulation. The time--averaged
torque $T_0=\Delta L/t$ that drives Neptune by increasing
its angular momentum by $\Delta L$ during the run--time $t$ is
shown in Figure \ref{torque}. This torque, as well as
the displacements $\Delta a$ of Fig.\ \ref{da}, should be regarded
as upper limits since resonance trapping did not occur in these
simulations. Had resonance capture been realized here,
an opposing torque on the planet would have developed
since the planet must transfer angular momentum to expand the orbits
of particles trapped at resonances. 

The back--torque due to a ring of mass $m$ that is captured at an
exterior $j+1:j$ mean--motion resonance can be calculated from the
rate of change of its angular momentum,
$L_m=m\sqrt{GM_\odot a(1-e^2)}$.  Since the resonant torque from
the planet expands the ring's semimajor axis at the rate
$\dot{a}/a=\dot{a}_p/a_p$ and also pumps up
eccentricities at the rate $de^2/dt=(\dot{a}_p/a_p)/(j+1)$
(\cite{RM93}), the rate of change of $L_m$ is
\begin{equation}
\frac{dL_m}{dt}\simeq\alpha\frac{m}{m_p} T
\end{equation}
where $\alpha=a_p/a=(j/(j+1))^{2/3}$ is the ratio of
the semimajor axes of the planet and the resonance site,
$m_p$ is the planet's mass, and
$T=m_p\sqrt{GM_\odot a_p}\dot{a}_p/2a_p$ is the net torque on the
planet, which simplifies to
$T = T_0-dL_m/dt \simeq T_0/(1+\alpha m/m_p)$.
Thus the back-torque from the resonance captured particles
will significantly slow the planet's orbital expansion if the mass
trapped at resonance is in
excess of the planet's mass: $m\gtrsim m_p/\alpha$.
In a generic system, the strongest exterior mean--motion
resonance is the 2:1 $(j=1, \alpha=0.63)$, which can be expected to
capture the most planetesimal mass.
If the capture efficiency is $\varepsilon$,
then we can estimate that planet migration will slow if
\begin{equation}
\varepsilon M_{D,ext}\gtrsim 1.6m_p
\end{equation}
where $M_{D,ext}$ is the mass of the disk exterior to the planet's
initial orbit.

If resonance capture is rather inefficient, then
a high--mass disk is simply fuel for planet--migration.
But if resonance trapping is effective, say, $\varepsilon\sim50\%$,
then the planet's migration will slow after the planet
has captured a mass $m$ comparable to its own at its exterior resonances.
In the disk models considered here (with
$M_{D,ext}\approx \twothirds M_D$), this would occur only in a
high--mass disk with $M_D\gtrsim80$ M$_\oplus$.

If migration by Neptune across a distance $\Delta a\sim5$--10 AU
is to explain the sculpted appearance of the Kuiper Belt, then
Figs.\ \ref{a(t)} and \ref{da} suggest that the initial planetesimal
disk must have had a mass $M_D$ in excess of $10$ M$_\oplus$
(since a lower--mass disk results in insufficient planet migration
that proceeds too slowly) but likely less than $\sim100$ M$_\oplus$
(since such a high--mass disk would likely produce additional giant
planets).  A more precise disk--mass estimate requires
detailed knowledge of the disk's radial extent and
the particulars of the disk surface density variations $\sigma(a)$.

\section{The Oort Cloud Mass}
\label{oort_cloud}

The observed flux of long--period comets
provides a constraint on the present mass of the Oort Cloud.
In this section, we use the results of our numerical simulations
to estimate the mass of its progenitor,
the primordial planetesimal disk in the outer Solar System.

After adjustment for the efficiency of comet detections
it is estimated that $\sim63$ long--period comets/year
pass through the inner Solar System with perihelia $q<4$ AU
(\cite{E67}).
This flux stems entirely from the Oort Cloud; objects
originating in the Kuiper Belt
evolve instead into the short--period Jupiter--family comets
(\cite{L96}). Dynamical models of the Oort Cloud, when
adjusted to match the flux of long--period comets,
require a reservoir of $N_{OC}\sim1.6\times10^{13}$
bodies\footnote{The \cite{H90} Oort Cloud model shows that a
reservoir of $10^{11}$ comets will produce a flux of 0.2
comets/year having $q<2$ AU. To match the observed flux
of comets with $q<4$ AU, multiply this population
by $63/(2\times0.2)$.} (\cite{H90}). Multiplying by the typical
comet mass yields the total mass of the Oort Cloud.
Based upon an admittedly uncertain relationship between
cometary brightnesses and their size, \cite{W(96)} concludes that the
mean comet mass is of order $\sim10^{16}$ gm, indicating an
Oort Cloud mass of order M$_{OC}\sim27$ M$_\oplus$.
However this mass estimate should be regarded as uncertain by at least
an order of magnitude since it relies upon a host of
uncertain quantities such as the efficiency of comet detections,
the comet brightness--mass relationship,
as well as uncertainties in the Oort Cloud perturbations (e.g., the
strength of the galactic tidal field, the frequency of stellar
encounters, {\it etc.}). 

The dynamics of the Oort Cloud is succinctly summarized by
\cite{DQT(87)}. Unless a particle is otherwise ejected from the system,
planetary perturbations cause its semimajor axis
to diffuse both inwards and outwards while keeping its
perihelion locked in the giant planet region of the Solar System.
However a more distant particle is susceptible to perturbations by the
galactic tide and passing stars
which cause its perihelion to diffuse on a timescale that
varies as $t_q\propto a^{-2}$. Those particles reaching
$3\times10^3\lesssim a\lesssim2\times10^4$ AU which have had their
perihelia raised well beyond the orbit of Neptune are thus decoupled
from the planets and are usually identified as {\sl inner} Oort Cloud
comets. However the perihelia of more distant bodies diffuse
more rapidly, and those with
$2\times10^4\lesssim a\lesssim10^5$ AU that reside
in the {\sl outer} Oort Cloud are in fact more likely to
diffuse back into the inner Solar System and
become observable as new long--period comets.
It is this flux of new comets that provides an important
constraint on the amount of mass driven from the initial disk
that has managed to avoid ejection during the last 4.5 Gyr.

Estimates of the Oort Cloud mass may be inferred from the data given
in Table \ref{oort_mass}. The quantity $f_{3k}$ is the fraction of the
{\sl dynamically active} disk that diffuses into the Oort Cloud
at $a>3\times10^3$ AU during each simulation.
The dynamically active disk is that part of
the disk where particles are likely to be perturbed into
Neptune--crossing orbits over the age of the Solar System. Long--term
integrations of Kuiper Belt orbits show that particles
in the active disk have semimajor axes
$a\lesssim1.4a_N$, where $a_N$ is Neptune's
semimajor axis (\cite{DLB95}). The mass of the dynamically
active disk, M$_{ad}$ (Table \ref{oort_mass}),
is defined here as the total mass of ejected
particles plus all survivors having $a\lesssim1.4a_N$
that are presumably in unstable orbits.
Also given is $f_h$, which is the fraction of the active
disk that has been ejected from the system.
Fig.\ \ref{fr} displays $f_{3k}$ and $f_{h}$ versus time
for the $M_D=100$ M$_\oplus$ run.
What is most striking is that at the end of all four runs
both $f_{3k}$ and $f_h$ vary little among the different simulations
(Table \ref{oort_mass}). This indicates that the total
mass deposited in the Cloud depends only on the mass
of the disk that lies within the planets' gravitational reach,
and is not very sensitive to the orbital histories of the
migrating planets. 

Ultimately, all particles starting in the dynamically active disk
are either ejected, deposited in the Oort Cloud, or in some instances
accreted by the planets. Although the latter outcome is not modeled
here, impacts may be assessed {\it ex post facto} using the collision
probabilities of \cite{O51}. After summing the probability
of each particle striking each planet, we find that $\sim15$ of the
1000 particles would have struck the giant planets during each
simulation, with roughly half of these impactors striking
Jupiter. Such impacts would have contributed no more than $\sim2\%$
to any planet's mass, so the neglect of particle--planet collisions is
justified. It is also worth noting that a few percent of the disk
passes through the terrestrial zone, as is
indicated by the $f_{q<3.5}$ curve of Fig.\ \ref{fr} which shows the
instantaneous disk fraction having perihelia inside of 3.5 AU. In
these simulations typically $\sim30\%$ of all disk particles have
brief episodes with $q<3.5$ AU. These findings are consistent with
earlier studies showing that planetesimals scattered during the epoch
of disk--clearing and planet--migration may have contributed
significant numbers of impactors during the late heavy bombardment
of the terrestrial planets (\cite{W75}, \cite{SW84}).

Table \ref{oort_mass} shows that in the four simulations the active
disks are depleted by a factor $f\equiv f_{3k}+f_h\simeq50\%$
after $t=3\times10^7$ years. It should be noted that the
family of $f(t)$ curves is described well by a power--law
$f\propto t^{0.44}$. This indicates that a fully evolved system having
$f\rightarrow1$ requires an integration lasting the duration of the
disk's dynamical lifetime $\tau_d\sim1.5\times10^8$ years,
which is beyond our computational means.
However it is straightforward to extrapolate the formation of the Oort
Cloud from the simulations at hand.
Fig.\ \ref{fr} shows that the ratio $f_{3k}/f_h\sim0.4$ remains
relatively constant during the bulk of the run.
This relation permits an extrapolation to a fully evolved state of the
system having $f_{3k}'+f_h'\rightarrow1$,
where the primes denote final extrapolated values. Assuming
$f_{3k}'/f_h'=f_{3k}/f_h$ over the age of the solar
system, the extrapolated fractions become
$f_h'=(1+f_{3k}/f_h)^{-1}\simeq0.73$ and $f_{3k}'=1-f_h'\simeq0.27$;
these values are also given in Table \ref{oort_mass}. Since these
disk fractions are both number as well as mass fractions,
the extrapolated planetesimal mass that is initially deposited in the
inner Oort Cloud is $f_{3k}'M_{ad}$.
However the galactic tide will subsequently strip away comets that
diffuse past $a\gtrsim10^5$ AU, and passing stars will eject others.
Numerical studies show that these external perturbations acting over
the age of the Solar System will reduce the Oort Cloud mass to about
a third of its initial value (\cite{DQT87}).
Thus the final Oort Cloud mass
reported in Table \ref{oort_mass} is $M_{OC}=f_{3k}'M_{da}/3$, and is
also displayed as the solid curve in Fig.\ \ref{Moc} as a function of
the initial disk mass $M_D$.

The nebula origin of the Oort Cloud is
given by the solid curve in Fig.\ \ref{OC_origin}, which
shows a histogram of the Oort Cloud particles' initial semimajor
axes for the $M_D=50$ M$_\oplus$ disk; after $t=3\times10^7$ years,
the planetary configuration in this system most resembles the Solar
System. Since the inner edge of our model disks is truncated at 10 AU,
any Oort Cloud mass originating in interior orbits is still
unaccounted for. We estimate this contribution after the fact by
evolving a system of four giant planets in their present configuration
plus an annulus of 153 massless test particles having an initial
$\sigma\propto a^{-1}$ surface number density between
$4<a<10$ AU. Their Oort Cloud contribution is given by the dashed
curves in Figs.\ \ref{Moc} and \ref{OC_origin}.
Evidently, all parts of the giant
planet domain contribute mass to the Oort Cloud. These
findings are in agreement with \cite{WL97} who reported
that a small fraction of Oort Cloud bodies can originate
from orbits interior to Jupiter and thus have asteroidal
rather than cometary compositions.

If the initial disk had a mass
$10\lesssim M_D \lesssim100$ M$_\oplus$ (Section \ref{migration}),
then Fig.\ \ref{Moc} indicates that the resulting Oort Cloud mass is
$0.5\lesssim M_{OC}\lesssim11$ M$_\oplus$.
This mass estimate must be qualified for two reasons.
For the lower--mass disks,
$M_D\le50$ M$_\oplus$, the mass estimate
$M_{OC}$ as given in Fig.\ \ref{Moc}
is likely an underestimate,
because in deriving it we have not accounted for the possibility
that Neptune can migrate deeper into the disk as the system
evolves further on timescales longer than $\sim50$ Myr,
thus allowing additional material to be injected into the
Oort Cloud.  (This does not affect the higher disk--mass simulations
since Neptune's gravitational reach
has already swept across the entire disk in the higher--mass runs.)
For the higher--mass disks,
$M_{OC}$ is likely overestimated in Fig.\ \ref{Moc}
because resonance trapping tends to slow planet--migration
and may reduce the mass encountering the planets and thus the mass
deposited in the Oort Cloud. 
We note that an Oort Cloud having a total mass $M_{OC}\sim5$ M$_\oplus$
and a population $N_{OC}\sim1.6\times10^{13}$ comets (\cite{H90})
suggests that a characteristic Oort Cloud comet has a radius
$\sim1$ km for a density $\sim0.5$ gm/cm$^3$.

\section{The Resonant, Stirred, and Scattered Kuiper Belt Components}
\label{scat_disk}

Although the giant planets scatter planetesimals
throughout the entire Solar System, a large gap in orbital phase
space will develop, as illustrated by Fig.\ \ref{disk}.
This gap is easily explained via the
restricted three--body problem, which shows that a planet on a
circular orbit will scatter a massless body in a manner that
preserves the particle's Jacobi integral $J$
(Eq.\ \ref{jacobi}). Particles originating in a cold disk in the
vicinity of the planet have semimajor axes $a\simeq a_p$
and $J\simeq3$. Once scattered, these particles will have
eccentricities defined by the curve
$e_{\mbox{\scriptsize J=3}}$ (Fig.\ \ref{disk}), assuming $i=0$.
It has also been shown that a planet of low 
eccentricity will repeatedly scatter particles
along a curve that approximately preserves $J\simeq3$
(\cite{IM93}). Thus when several planets are present, those interior
to Neptune can loft particles into high--eccentricity
Neptune--crossing orbits and ultimately fill the
$e>e_{\mbox{\scriptsize J=3}}$ region of phase space.
Those particles in the high--eccentricity 
$e\gtrsim e_{\mbox{\scriptsize J=3}}$ orbits of Fig.\ \ref{disk}
are referred to as the scattered disk (e.g., \cite{DL97}).
The region occupied by bodies of lower
eccentricity beyond $\sim35$ AU is sometimes referred to in
recent literature as the `classical' Kuiper Belt;
the eccentric, resonant objects (including the so-called
Plutinos at Neptune's 3:2 resonance) may be considered a
distinct dynamical sub-class of the latter population.

Although Fig.\ \ref{disk} shows the scattered and classical disks
after only $5\times10^7$ years, \cite{DL(97)} have integrated
test particle orbits for the age of the Solar System.
They find that while most particles in the scattered disk
are removed in less than the age of the Solar System,
about $1\%$ of the scattered particles survive
for longer times (perhaps by acquiring protection via the Kozai
mechanism (\cite{K62}) or by sticking near mean--motion resonances).
In the
absence of disk--stirring by any other large distant perturbers,
we note that the
phase--space gap between the classical and
scattered disk will persist over the age of the Solar System.
This gap should become evident as deeper
observations begin to peer beyond Neptune's 2:1 resonance.

Until very recently, no KBOs were known to orbit at Neptune's 2:1
resonance,  in apparent conflict with the prediction of the
planet--migration/resonance--sweeping theory that KBO populations
at the 2:1 and 3:2 should be similarly abundant and have comparable
eccentricities (\cite{RM95}).
As this paper was being written, we learned that two KBOs,
1997 SZ$_{10}$ and 1996 TR$_{66}$,
have been identified as librating in the 2:1 Neptune resonance
(\cite{BM98}).
We note that these two KBOs were formerly identified in Neptune's
5:3 and 3:2 resonances, and that their orbits were revised to
the 2:1 resonance only after observations spanning two and three
oppositions, respectively.
Clearly, orbit-fitting biases and observational incompleteness
remain in the current census of the Kuiper Belt.
A robust test of the planet--migration/resonance--sweeping theory
requires a larger observational sample of reliable KBO orbits.

\cite{MV(97)} offer an alternative explanation for the structure
in the Kuiper Belt.
They suggest that if a stationary Neptune had 
scattered a couple of Earth--mass planetesimals
(e.g., LNSPs = large Neptune--scattered planetesimals)
outwards, these massive bodies could have
stirred up KBO eccentricities to $e\sim0.2$, similar to those
observed for KBOs at Neptune's 3:2 resonance.
Gravitational scattering can indeed insert KBOs into, as well as
remove objects from, mean--motion resonances, but scattered objects
tend to librate at resonance with such large amplitudes that close
encounters with Neptune become possible and long--term orbital
stability is precluded (\cite{LS95}). Therefore an additional sequence
of collisions and/or scattering events is
required in order for particles to diffuse to stable, low--amplitude
librating orbits. Only the fortunate few would survive this process,
so the yield of KBOs scattered into stable $e\sim0.2$
orbits at the 3:2 resonance would be much smaller than that which
might otherwise be acquired by means of adiabatic orbit expansion and
resonance--sweeping by Neptune.
\cite{MV(97)} also point to the $e\sim0.1$ KBOs that reside between
the 3:2 and 2:1 resonances as additional evidence for LNSPs.
But resonance capture is not entirely efficient, and similar
eccentricities are also achieved as Neptune's orbit expands and
its 2:1 resonance sweeps across the disk and stirs up
eccentricities (Fig.\ \ref{disk}).

A census of KBOs beyond
Neptune's 2:1 resonance (at 48 AU) would permit an evaluation
of possible stirring by hypothetical LNSPs. If the natal planetesimal
disk does extend past 48 AU, then Fig.\ \ref{disk} shows that
planet migration will produce a stirred zone interior to the 2:1
yet leave the disk exterior to the 2:1 relatively undisturbed.
However, an abundant population of eccentric KBOs beyond the 2:1 would
suggest a history of additional stirring by other unseen perturbers
(though this would not preclude an episode of planet--migration).

\section{Disk Self--Gravity and the Role of Spiral Density Waves}
\label{waves}

As noted earlier, our simulations neglect the planetesimal
disk's self--gravity in order to inhibit an unphysical degree of
self--stirring. The consequences of disk self--gravity could
be better studied only by simulating disks composed of many more
lower--mass particles. In this section, we discuss the possible
consequences of the disk's self--gravity.

If the local disk mass exceeds the mass of a nearby
planet (as is the case for Neptune in
most of the simulations considered here),
it is the disk's gravity that can control
the rates at which the perihelia and nodes of both
the planet and the disk particles precess and/or regress (\cite{W81}).
Secular resonances are sites in the disk where a planet's
perihelion/node varies at the same rate as the disk particles',
and large eccentricities and inclinations can get excited at these
resonances. As planets sculpt the disk
and cause its surface density to evolve over time, the location
and strength of secular resonances will shift.
Although radial drifts in the location of secular resonances might
alter the details of how a planet depletes the disk as it excites
particles into crossing orbits,  this issue is likely of lesser
importance when compared to the mean--motion resonances.

In particular, a planet that is embedded in a self--gravitating disk
can launch spiral density waves at its mean--motion
resonances. Numerous examples of this
phenomenon exist in Saturn's rings which exhibit density waves driven
by orbiting satellites. The gaps in these rings reveal a history of
angular momentum exchange between ring material and satellites, and
similar exchanges are expected of planet--forming systems.
When Neptune launches density waves at an exterior $j+1:j$
mean--motion resonance, the disk exerts the torque $T_j$
on the planet which opposes its radial migration (\cite{GT80}):
\begin{equation}
\label{resonant_torque}
T_j=\frac{j\pi^2\sigma\psi^2}{rdD/dr}\simeq
-\frac{64\pi j^3}{75(j+1)}\mu_p^2\mu_dM_\odot(a_p\Omega_p)^2.
\end{equation}
Here, 
$\psi\simeq-8j\mu_p(r\Omega)^2/5$ is the planet's forcing function,
$\mu_p$ is the planet's mass in solar units, $\Omega$ is the disk's
mean motion, $D$ is the frequency difference from exact resonance
which has the gradient $rdD/dr\simeq-3(j+1)\Omega^2$, and the preceding
quantities are evaluated at resonance $r=(1+j^{-1})^{2/3}a_p$
where $a_p$ and $\Omega_p$ are the planet's semimajor axis 
and mean motion (\cite{HWR95}). The dimensionless `disk mass' is
$\mu_d\equiv\pi\sigma(a_p)a_p^2/M_\odot\simeq
1.1\times10^{-6}(M_D/M_\oplus)$ for the disk simulations that have
$\sigma(a_p)=1.3\times10^{-4}$ M$_D$/AU$^2$.
Figure \ref{torque} sums the torques $T_j$ due to Neptune's
outer $j=1$ to 5 resonances for comparison with the non--resonant
torque $T_0$ that drives Neptune's orbit expansion.

Evidently, the resonant disk torques can inhibit
Neptune's outward migration if the disk admits a wave response
at the $j\gtrsim3$ resonances. It should be
noted that these torques are also operative in non--self--gravitating
disks (\cite{LE98}), and similar torques also slow orbit expansion when
resonance trapping is effective (Section \ref{back_torque}).
But when trapping is not of concern, the resonant torques
in a non--gravitating disk shut off once particle
eccentricities get excited (\cite{LE98}).
However in a self--gravitating particle disk, density waves transport
the planet's forced disturbance downstream of the resonance in the
direction of the planet's orbit. As long as wave action is
sustained, particles at resonance maintain low eccentricities and the
resonant disk torque can oppose planet--migration.

There are two ways in which the propagation
of density waves might be inhibited in
particle disks. In a disk that is populated by comet--sized (or larger)
planetesimals, dissipative forces such as gas drag or
viscosity due to inter-particle collisions are insufficient to
damp out density waves (e.g., \cite{HWR95}). In this case,
density waves reflect at a $Q$--barrier in the disk and return to the
launch zone where they are reabsorbed by the particles at resonance
(\cite{T69}). This absorption of the returning waves' energy
will steadily heat the disk and can eventually shut off subsequent
wave generation. A second way to defeat waves is via 
stirring by larger bodies which also heats the disk and inhibits wave
propagation. Consequently, the resonant torque that the particle
disk exerts on Neptune will delay the onset of orbit expansion
until the disk becomes too stirred to sustain density waves at its
$j\gtrsim3$ resonances.

\section{Conclusions}
\label{conclusions}

The existence of the Oort Comet Cloud as well as the Kuiper Belt
suggest that there was once a residual planetesimal disk of mass
$\sim10$--100 M$_\oplus$ in the vicinity of the giant planets
following their formation. Further, any inefficiencies in the
formation of the giant planets' cores implies additional disk mass.
The eventual clearing of this planetesimal population involves
a substantial exchange of orbital energy and angular momentum with
the planets, implying that
the present locations of the giant planets are not necessarily
their formative ones.
We have numerically simulated the evolution of a system of
four giant planets embedded in a planetesimal disk of mass
ranging from $10M_\oplus$ to $200 M_\oplus$.
Our numerical simulations show a gradual increase in the
mutual separation of the planets' orbits as the disk is
dispersed via gravitational scattering by the planets.
Higher disk masses yield planetary orbital migration that is
faster and larger in magnitude.
If planet--migration
and resonance--trapping is invoked to explain the eccentricities of
Pluto and its cohort of Kuiper Belt Objects at Neptune's 3:2
mean--motion resonance, then these simulations show that
a disk mass of order $M_D\sim50$ M$_\oplus$ is required
to expand Neptune's orbit the requisite
distance of $\Delta a\sim7$ AU to pump up Plutino
eccentricities to $e\sim0.3$. Such an
episode of planet migration implies that the initial Solar System
was more compact in the past, with the Jupiter--Neptune separation
having been smaller by about $30\%$. This finding also confirms the
disk mass estimate previously obtained by \cite{RM97}.

Our model disk--planet systems behave similarly
to other disk systems that
experience a gravitational or viscous torque (e.g., \cite{LK72},
\cite{LP74}), which causes angular momentum to be carried radially
outwards (in this application, by the outer three planets) while disk
particles deliver mass radially inwards. However these particles
tend to get ejected upon reaching Jupiter's orbit, which accounts for
that planet's slight orbital decay.

Since our simulations neglected the disk's self--gravity,
collective effects such as density waves are precluded.
A planet embedded in a self--gravitating planetesimal disk will tend to
launch spiral density waves at its resonances. The torque
due to wave generation is sufficient to oppose Neptune's orbit
expansion as long as the disk remains dynamically cold enough to admit a
wave response from its $j\simeq3$ or higher resonances.
Such an episode of wave generation will delay the onset of planet
migration until the disk's wave response is defeated.

The bulk of the disk particles deposited in the
Oort Cloud originate in the vicinity of the Saturn--Neptune region of
the solar nebula. Assuming that galactic tides
and passing stars decouple particles from the planetary system when
they achieve a semimajor axis of $a>3000$ AU, and that these
perturbations also remove about two--thirds of the Oort Cloud over the
age of the Solar System (\cite{DQT87}), we estimate that
about $M_{OC}\sim12$ M$_\oplus$ of the $M_D=50$ M$_\oplus$
disk is initially emplaced in the Oort Cloud,
of which $\sim4$ M$_\oplus$ will persist to the present age of the
Solar System.

Due to the fact that the disks simulated here were sparsely populated
by particles having masses $m=0.01$--0.2 M$_\oplus$, their vigorous
scattering caused the planets' orbits to evolve non--adiabatically such
that resonance trapping of KBOs was inhibited. However previous
studies have shown that adiabatic orbit
expansion by Neptune can account for the abundance of eccentric KBOs
that are known to orbit at Neptune's 4:3, 3:2, 5:3, and 2:1
mean--motion resonances (\cite{RM95}). Unless there are (or were)
additional unseen, distant perturbers, any primordial KBOs 
beyond Neptune's 2:1 resonance should reside in
nearly circular, low--inclination orbits.

The planet--migration/resonance trapping phenomenon might also have
applications in extrasolar planetary systems. The most visible
component of an extrasolar planetary system is likely its dust,
which should be most abundant when planets and planetesimals are
colliding, accreting, and eroding. Dusty circumstellar disks and
rings are known
to orbit the stars $\beta$ Pictoris, Formalhaut, HR 4796A,
55 Cancri, and $\epsilon$ Eridani (\cite{ST84}, \cite{Greaves98},
\cite{TB98}, \cite{Ketal98}, \cite{Hetal98}, \cite{Setal98}).
In some of these systems, collisions and/or radiation forces will
remove the observed dust on a timescale shorter than the age
of the parent star. The presence of dust
thus suggests an additional source---perhaps dust generation due to
collisions by unseen planetesimals that reside in the disk.
Any planets that might form within
this environment will deplete the disk region that lies within
their gravitational reach, which could account for these disks'
central gaps. However an episode of planetesimal disk--clearing
would also drive planet--migration, which can concentrate
planetesimals at the
outermost planet's exterior mean--motion resonances. Since the
collision frequency and hence the dust generation rate varies as the
square of the planetesimal density, one might speculate that this
mechanism is also responsible for the formation of dust rings
observed around $\epsilon$ Eridani, Formalhaut, and HR 4796A.

\acknowledgments
\begin{center}
{\bf Acknowledgments}
\end{center}

The authors thank Derek Richardson for a careful
review of this paper.
This paper is contribution 966 from the Lunar and Planetary
Institute, which is operated by the Universities Space Research
Association under NASA contract NASW--4574. This research was
supported in part by NASA's Origins of Solar Systems Research Program.


\clearpage
\begin{table}
\caption{\label{oort_mass}}
\begin{tabular}{ccccccc}
\multicolumn{7}{c}{\bf TABLE \thetable}\\
\multicolumn{7}{c}{\bf Oort Cloud Masses}\\
\hline\hline
$M_D$		& $f_{3k}$	& $f_h$	& $f_{3k}'$	& $f_h'$	& M$_{ad}$	& 	$M_{OC}$	\\
(M$_\oplus)$	& 		&	&		& 		& (M$_\oplus$)  	& (M$_\oplus$)	\\
\hline
10		& 0.097		& 0.355	& 0.215		& 0.785		& 	5.7		& 0.41		\\
50		& 0.143		& 0.364	& 0.282		& 0.718		& 	35		& 3.3		\\
100		& 0.142		& 0.327	& 0.303		& 0.697		& 	99		& 10		\\
200		& 0.141		& 0.369	& 0.276		& 0.724		& 	200		& 18		\\
\hline
\end{tabular}
\end{table}

\clearpage

\begin{figure}
\caption{Fractional variations
in a particle's Jacobi integral $J$ as a function of a particle's
perihelion distance $q$ averaged over $10^4$ years. Jupiter lies on a
circular orbit at 5 AU and the integrator step size is $\Delta t=0.4$
years. Squares indicate particles having initial eccentricities of
$e=0.1$, inclinations $i=3^\circ$, and
semimajor axes $4<a<50$ AU, while crosses and circles
are for eccentric particles having initial perihelia $0.05<q<6$ AU,
$a=6$ AU, and $0<i<10^\circ$.
Filled circles and squares indicate particles that approached
closer than a Hill radius, or 0.35 AU, of Jupiter.}
\label{delta J}
\end{figure}

\begin{figure}
\caption{A system consisting of Neptune on a circular orbit
at $a=30$ AU and 100 massless particles having initial
semimajor axes $28.5<a<31.5$ AU, eccentricities $e=0.05$,
inclinations $i=0.025$ radians, and Jacobi
integrals $2.997<J<2.999$ is evolved for $5\times10^7$ years.
The solid curve gives the swarm's average
$J$ with vertical bars indicating the swarm's standard
deviation $\sigma_J$. The characteristic timescale to diffuse into
Uranus--crossing orbits having $J<J_U$ that are dynamically forbidden
is $t_d\sim2.5\times10^7$ years.}
\label{J diffusion}
\end{figure}

\begin{figure}
\caption{(a) The fractional error in the relative velocities
$\delta V/V$ versus periapse $q$ (in units of Neptune's Hill radius
$R_H$) of 500 particles after scattering off Neptune. The particles'
initial heliocentric orbits were $a\simeq30$ AU, $e=0.05$, and
$i=0.025$ radians. Particles that trigger the two--body close encounter
algorithm are indicated by a dot. (b) A histogram of the
$z$--component of the specific angular momentum errors
$\delta\ell_p$ for each scattering event. The total specific angular
momentum exchanged between the planet and the particle swarm is
$L=|\sum\ell_p|=1.27$ AU$^2$/yr which has an rms sum
$(\sum\ell_p^2)^{1/2}=0.64$ AU$^2$/yr and an rms error
$\delta L=0.020$ AU$^2$/yr .}
\label{dV}
\end{figure}

\begin{figure}
\caption{The semimajor axes of the the giant
planets while embedded in planetesimal disks of mass $M_D=10$, 50,
100, and 200 M$_\oplus$. The boundaries of the grey regions denote the
planets' perihelia and aphelia distances.}
\label{a(t)}
\end{figure}

\begin{figure}
\caption{The eccentricities $e$ versus semimajor axes $a$ at
logarithmic time intervals for the M$_D=50$ M$_\oplus$ system.
Small dots indicate scattered particles that have
passed within a Hill radius of a planet and
crosses indicate particles that have not encountered a planet.
Large dots denote the planets and the vertical dashes
indicate the location of Neptune's outer four mean--motion
resonances. Orbits lying above the left curve have perihelia inside
of Saturn's orbit and those above the right curve have perihelia
interior to Neptune.}
\label{ea}
\end{figure}

\begin{figure}
\caption{The simulation of the
$M_D=50$ M$_\oplus$ system extended out to $t=5\times10^7$ years.
Grey indicates perihelia and aphelia distances.
Note that Uranus and Neptune pass through a 2:1 mean--motion resonance
at $t=3.05\times10^7$ years, which results in brief eccentricity
excitation.}
\label{a(t)_50}
\end{figure}

\begin{figure}
\caption{The radial displacement $\Delta a$
versus disk mass $M_D$ for each planet after $t=3\times10^7$ years.}
\label{da}
\end{figure}

\begin{figure}
\caption{The solid curve is the
time--averaged torque $T_0$ on Neptune, in units of that planet's
current angular momentum/orbital period ratio $L/P$, and is plotted
versus the disk mass $M_D$.
The dotted curves indicate sums of the resonant disk torque
contributions, Eq.\ \ref{resonant_torque}.}
\label{torque}
\end{figure}

\begin{figure}
\caption{The fraction of the active disk that
is ejected into hyperbolic orbits, $f_h$, versus time $t$ for the
$M_D=100$ M$_\oplus$ simulation. The disk fraction
that is scattered into the Oort Cloud with $a>3000$ AU is $f_{3k}$, and
$f_{q<3.5}$ is the instantaneous disk fraction having perihelia
$q<3.5$ AU. Note the near constancy of the ratio
$f_{3k}/f_h\simeq0.4$.  Curves for the other runs are quite similar.}
\label{fr}
\end{figure}

\begin{figure}
\caption{The extrapolated mass of the Oort Cloud
mass $M_{OC}$ as a function of the initial disk mass $M_D$. The dashed
curve includes contributions by massless
test particles originating in the $4<a<10$ AU part of the disk that is
modeled separately. However each one of these particles, when
deposited in the Oort Cloud, are assumed to contribute the same
individual masses as their compatriot particles that are employed
in the simulations of Fig.\ \ref{a(t)}.}
\label{Moc}
\end{figure}

\begin{figure}
\caption{The local
disk fraction deposited in the Oort Cloud versus initial semimajor
axis $a$ for the $M_D=50$ M$_\oplus$ simulation.
The data are obtained at time $t=3\times10^7$ years
and then extrapolated to a fully evolved state by multiplying by
$f'_{3k}/f_{3k}=1.97$. The grey bars show the extent of
radial migration by each planet. The $4<a<10$ AU component
(dashed curves) is obtained from a separate integration,
and $N^{1/2}$ errors are assumed.}
\label{OC_origin}
\end{figure}

\begin{figure}
\caption{Eccentricity $e$ versus
semimajor axis $a$ at time $t=5\times10^7$ years for the $M_D=50$
M$_\oplus$ simulation. Large dots indicate Uranus and Neptune,
small dots indicate scattered particles that have passed within a Hill
distance of a planet, and crosses denote particles that have not had a
close planetary encounter. The dashed lines indicate
Neptune's four outermost mean--motion resonances, and the
$e_{\mbox{\scriptsize J=3}}$ curve satisfies
Eq.\ (\ref{jacobi}) with $i=0$. Boxes denote observed KBOs
having well--determined orbits (from Marsden (1998)), and scattered
object 1996 TL$_{66}$ is indicated.}
\label{disk}
\end{figure}


\clearpage
\begin{figure}[t]
\psfig{figure=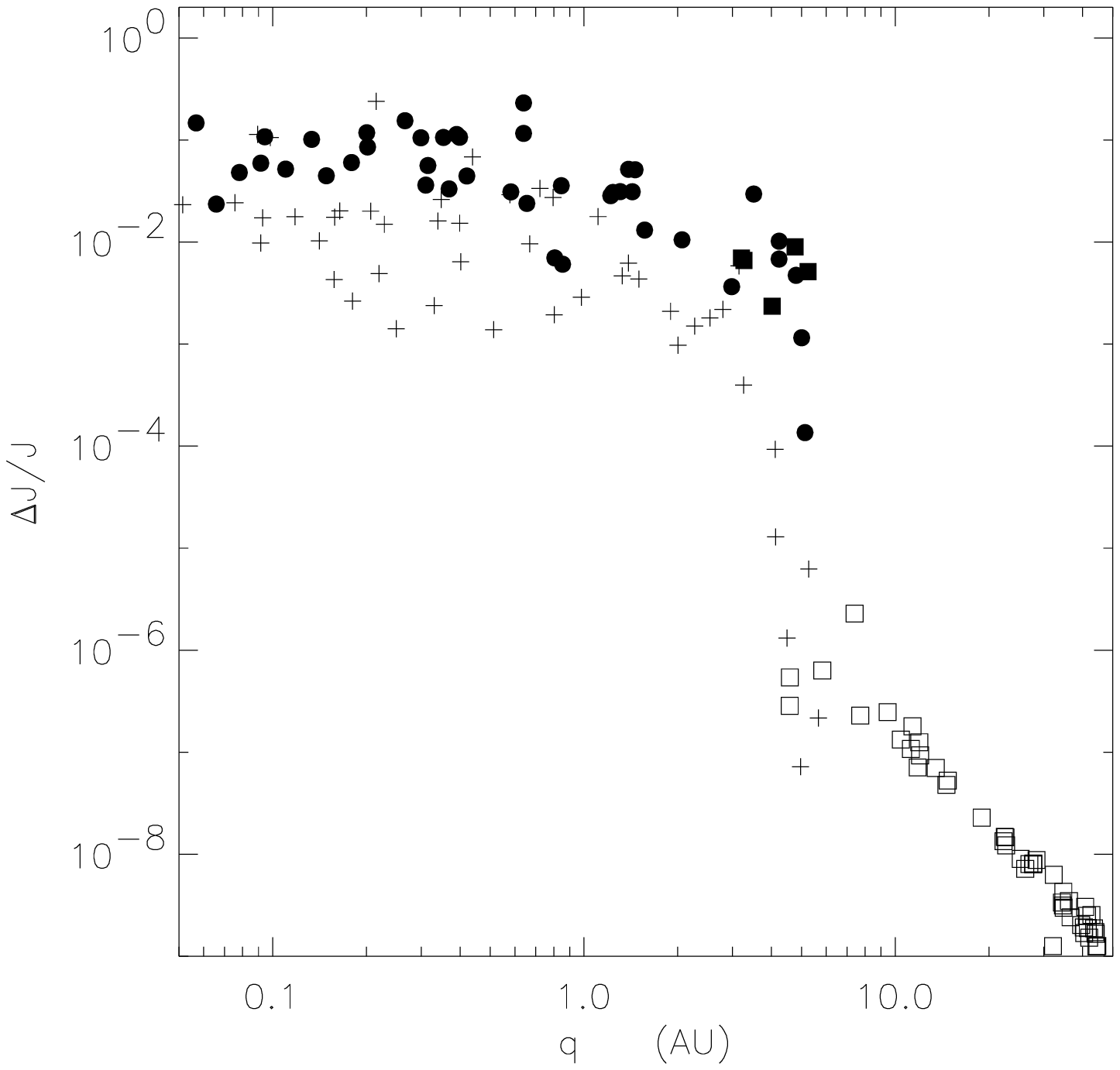,height=6.0in,width=6.0in}
Figure \ref{delta J}.
\end{figure}

\clearpage
\begin{figure}[t]
\psfig{figure=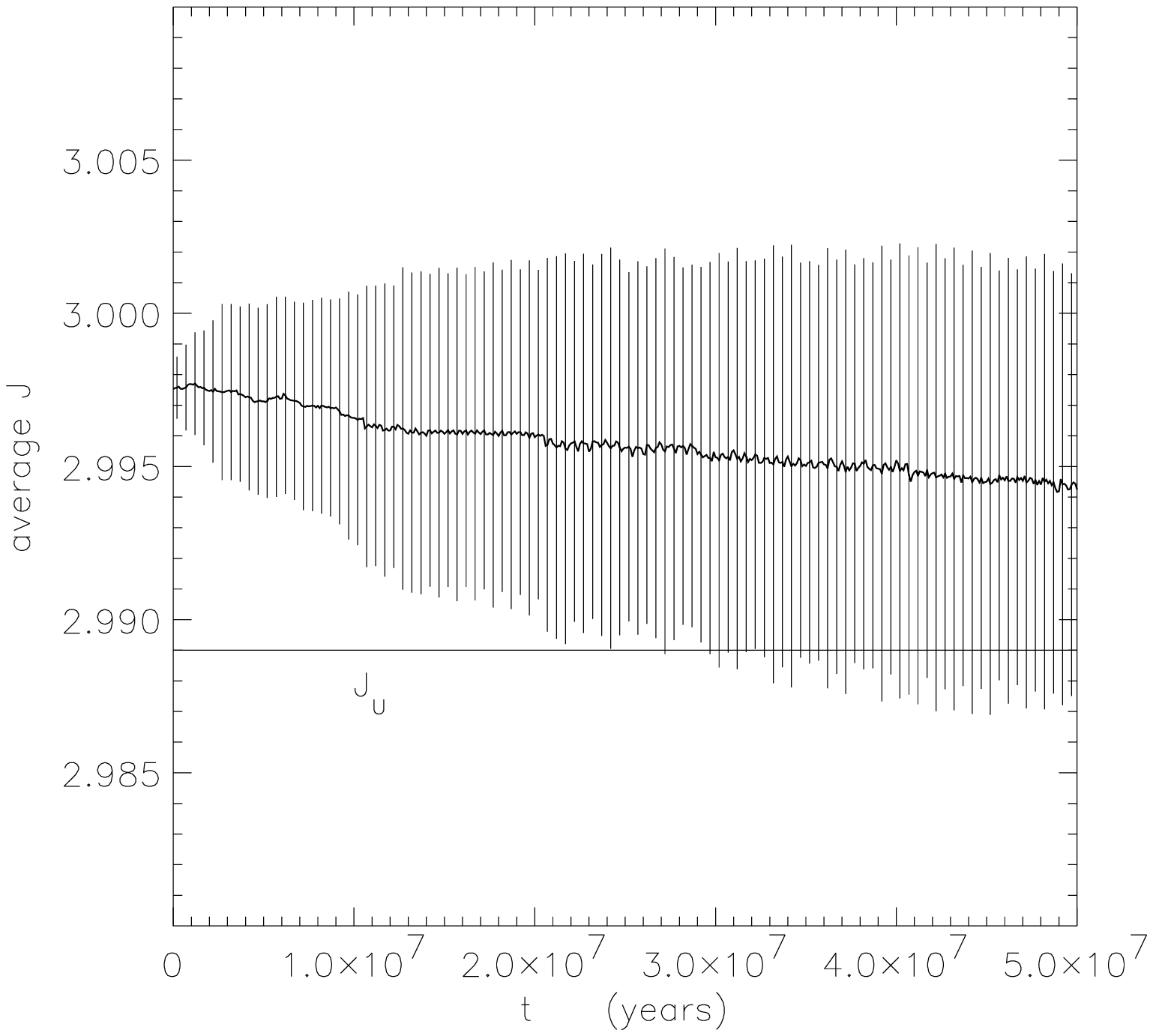,height=6.0in,width=6.0in}
Figure \ref{J diffusion}.
\end{figure}

\clearpage
\begin{figure}[t]
\psfig{figure=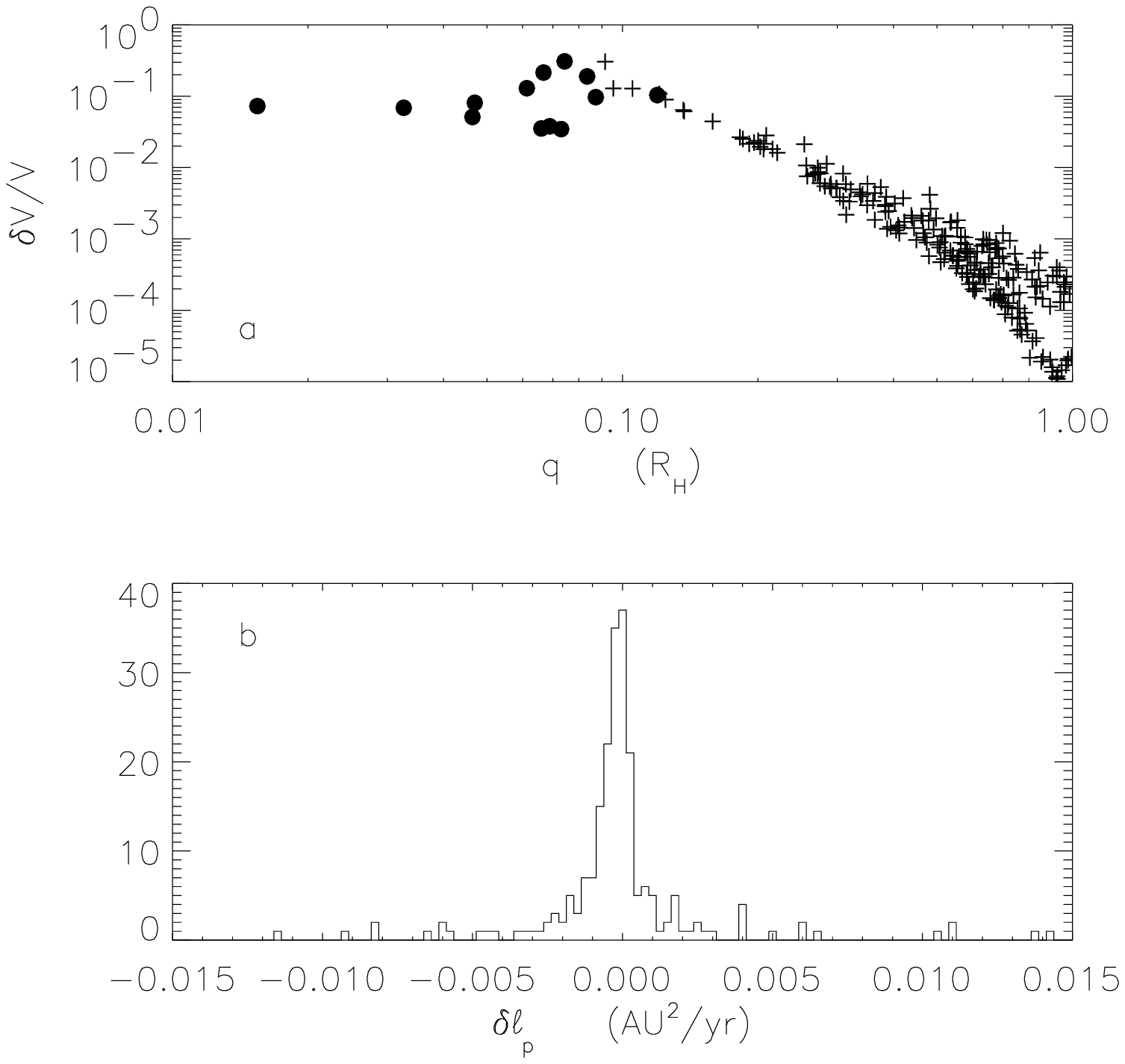,height=6.0in,width=6.0in}
Figure \ref{dV}.
\end{figure}

\clearpage
\begin{figure}[t]
\centerline{\psfig{figure=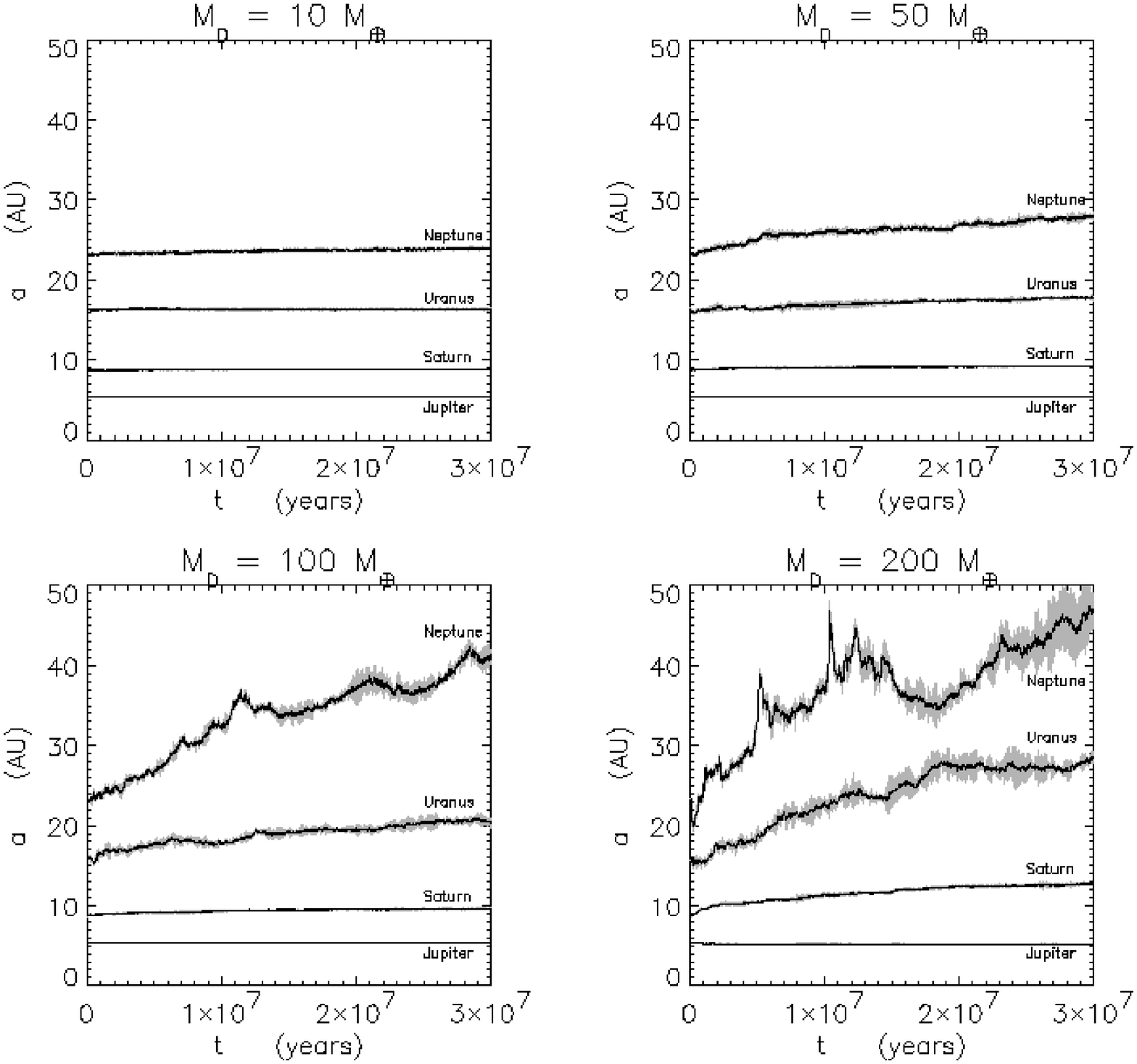,height=7.5in,width=7.5in}}
Figure \ref{a(t)}.
\end{figure}

\clearpage
\begin{figure}[t]
\psfig{figure=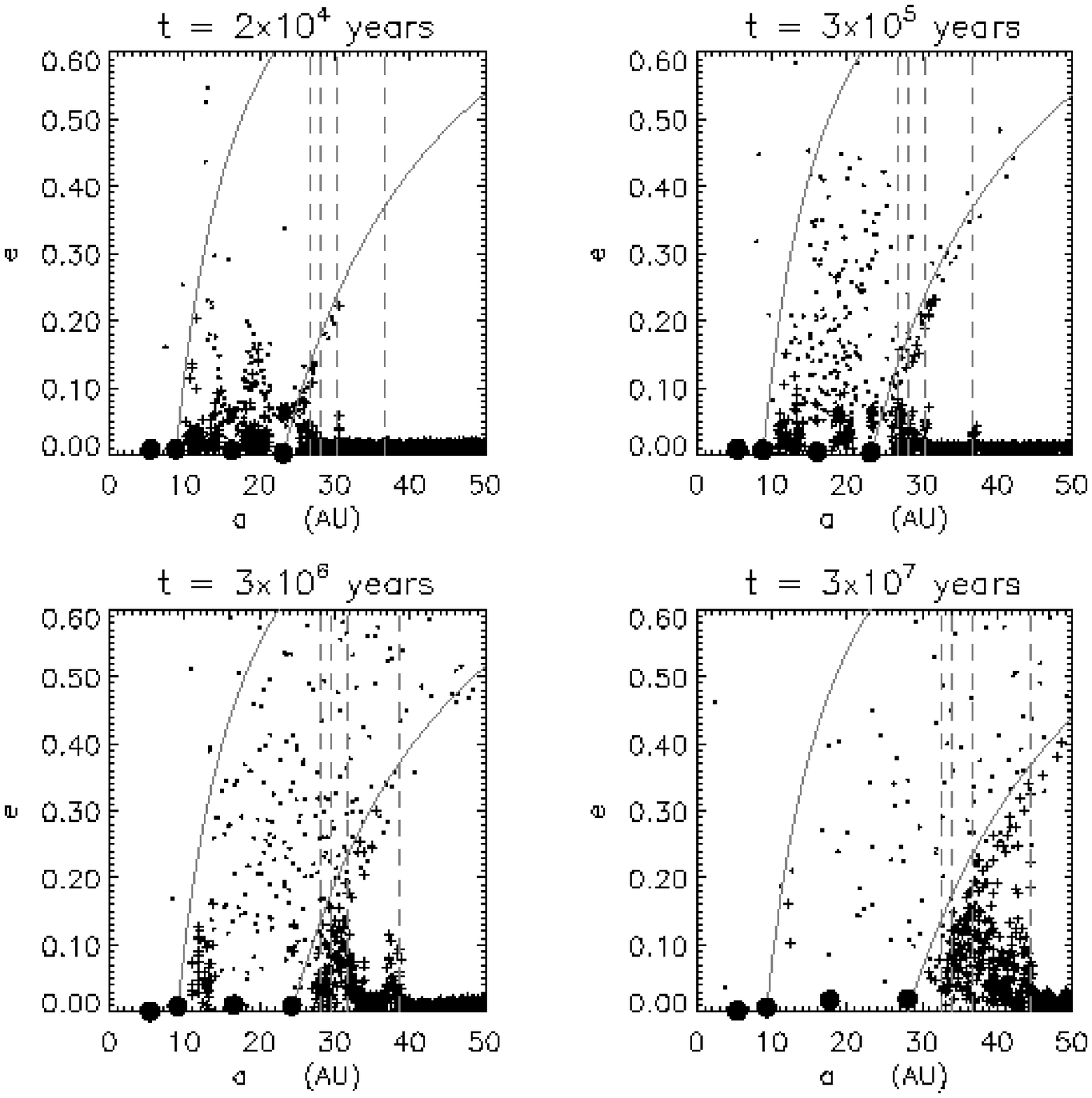,height=7.0in,width=7.0in}
Figure \ref{ea}.
\end{figure}

\clearpage
\begin{figure}[t]
\psfig{figure=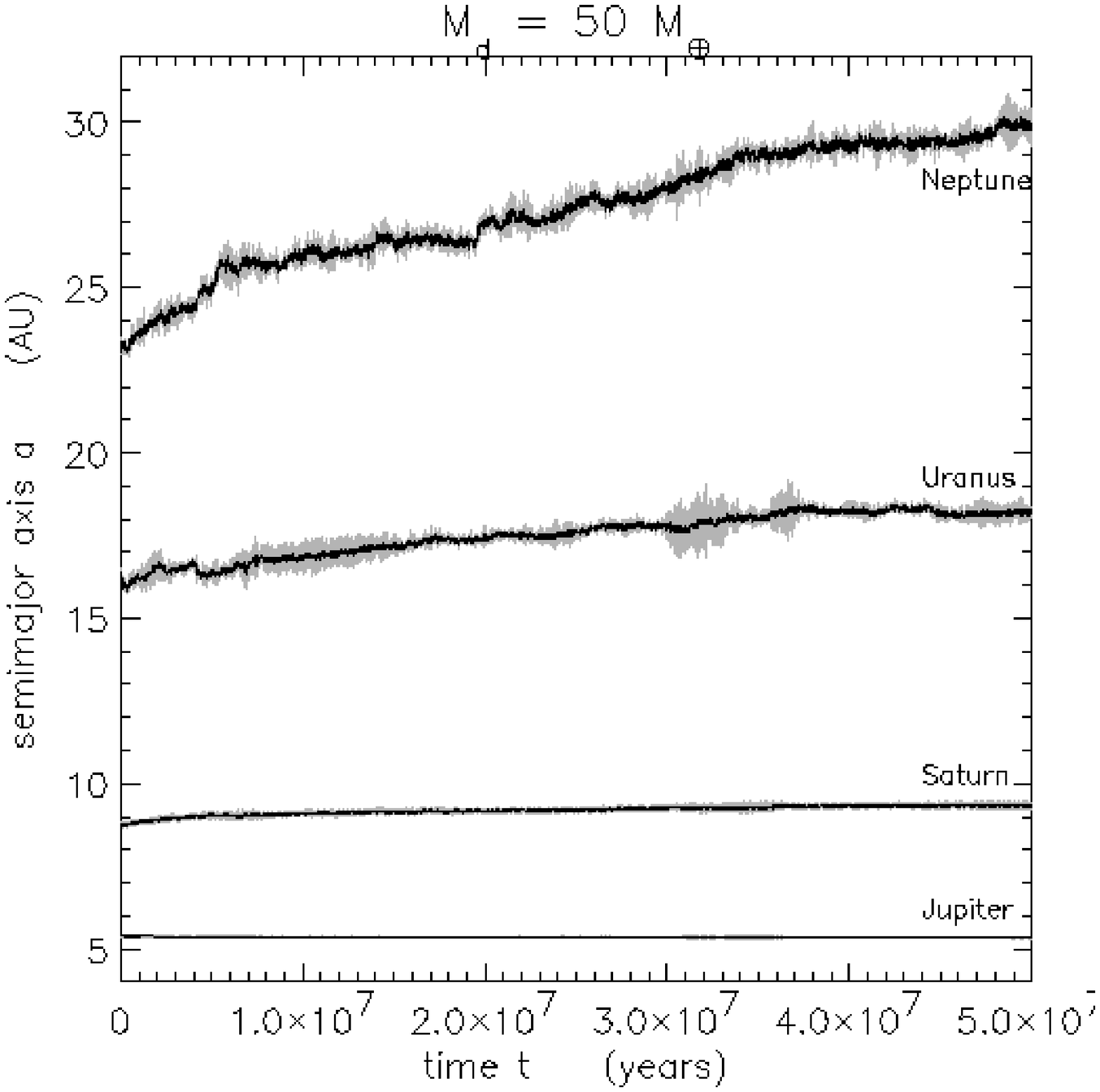,height=6.0in,width=6.0in}
Figure \ref{a(t)_50}
\end{figure}

\clearpage
\begin{figure}[t]
\psfig{figure=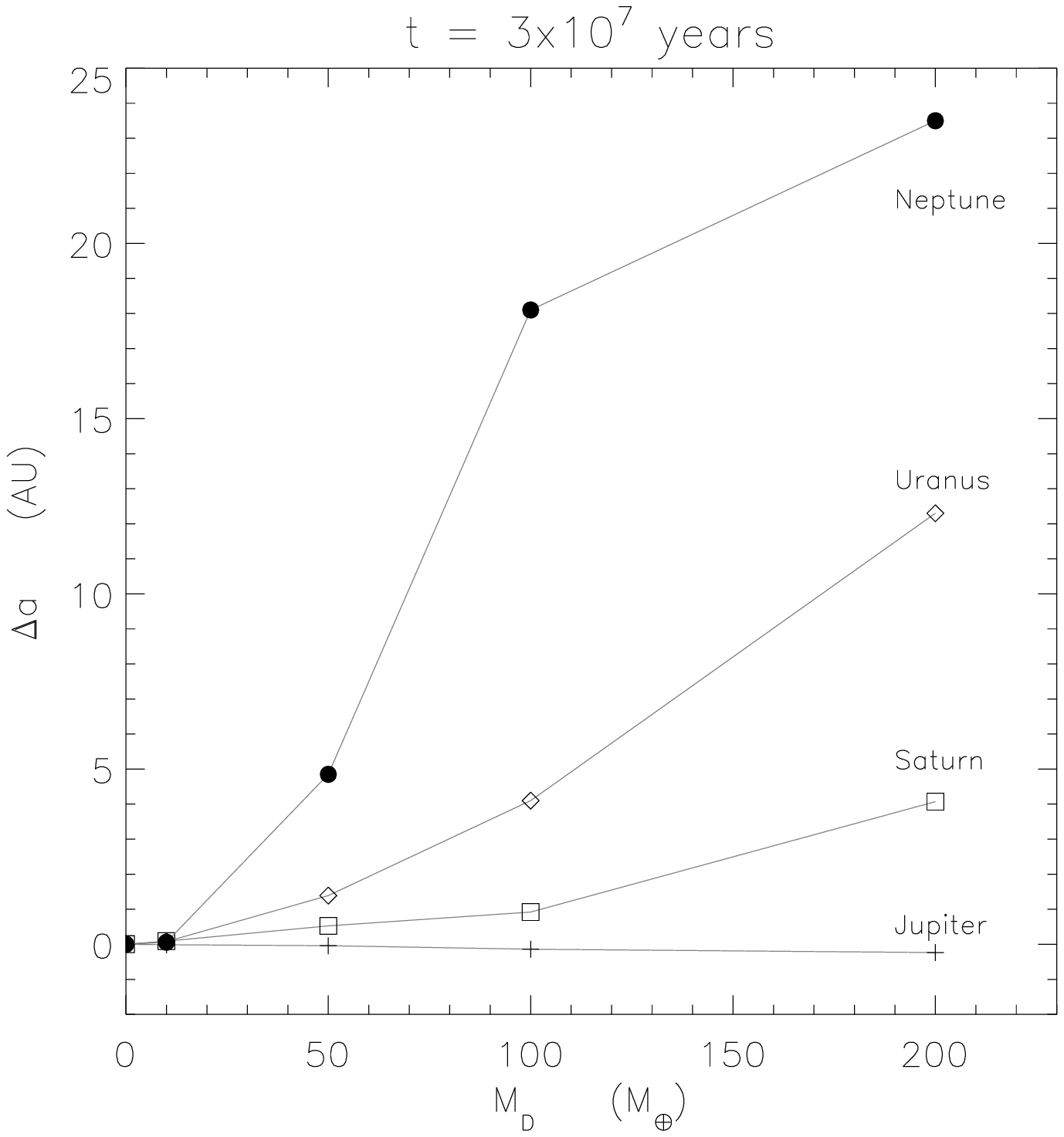,height=6.0in,width=6.0in}
Figure \ref{da}
\end{figure}

\clearpage
\begin{figure}[t]
\psfig{figure=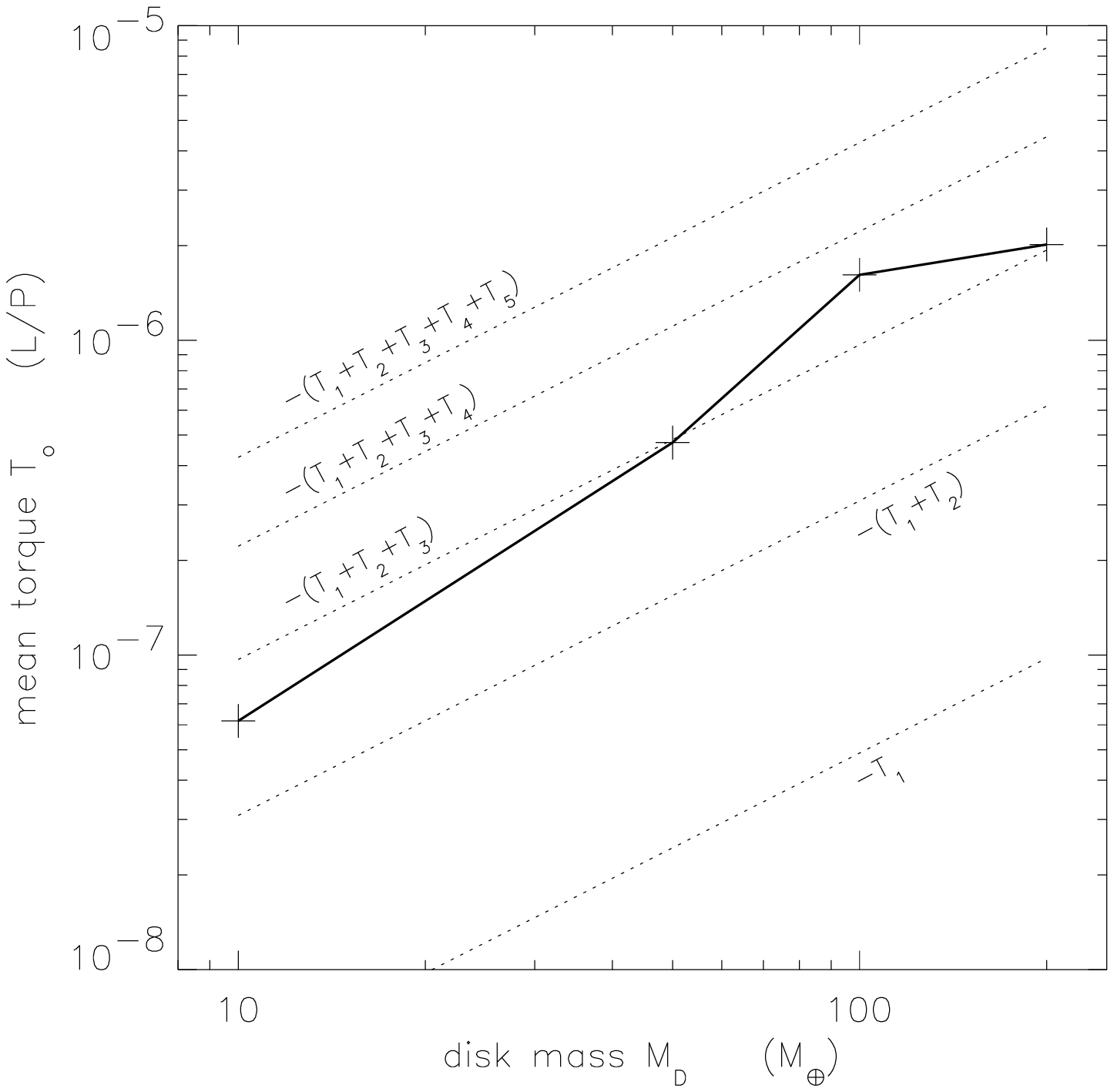,height=6.0in,width=6.0in}
Figure \ref{torque}
\end{figure}

\clearpage
\begin{figure}[t]
\psfig{figure=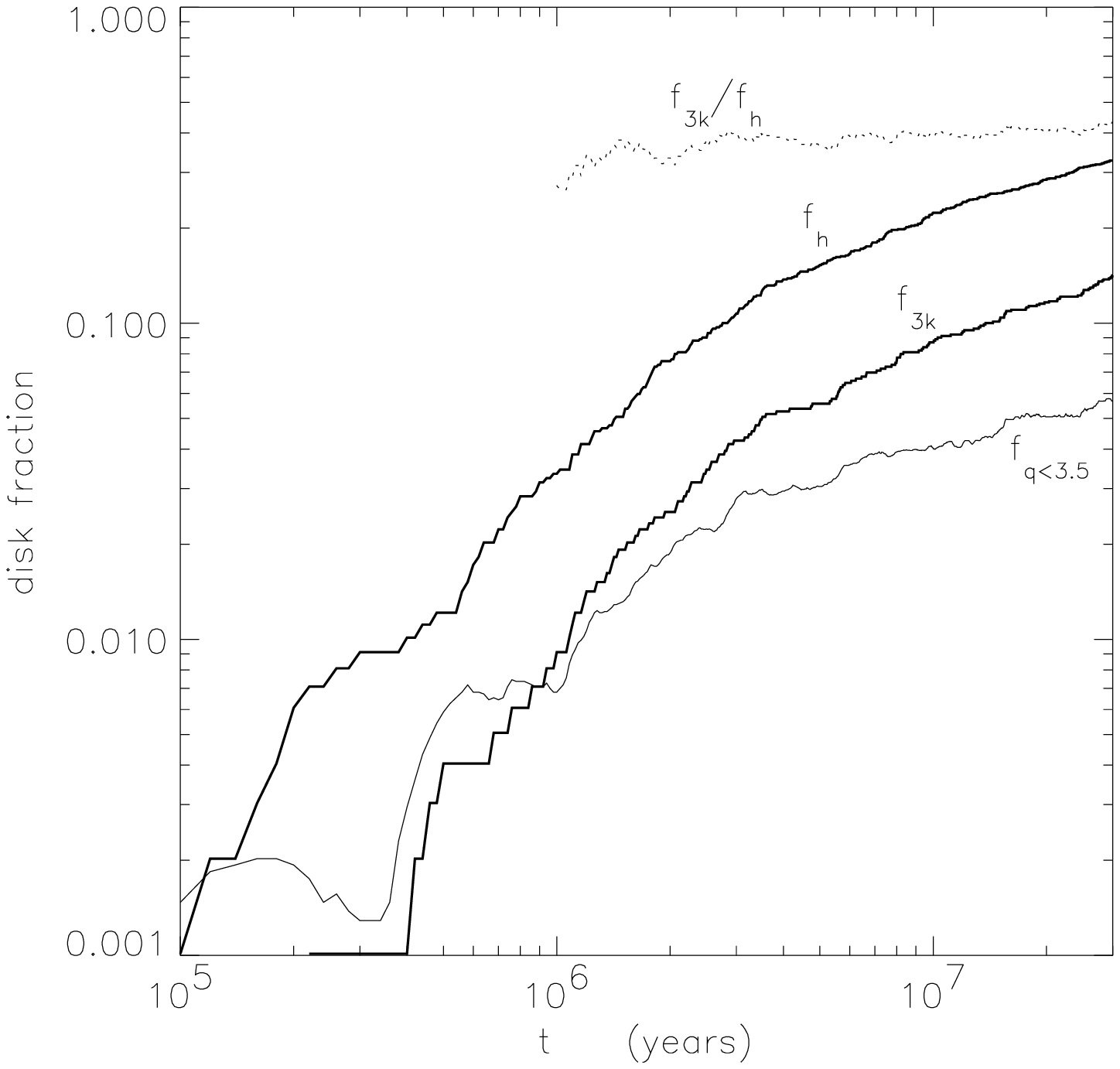,height=6.0in,width=6.0in}
Figure \ref{fr}
\end{figure}

\clearpage
\begin{figure}[t]
\psfig{figure=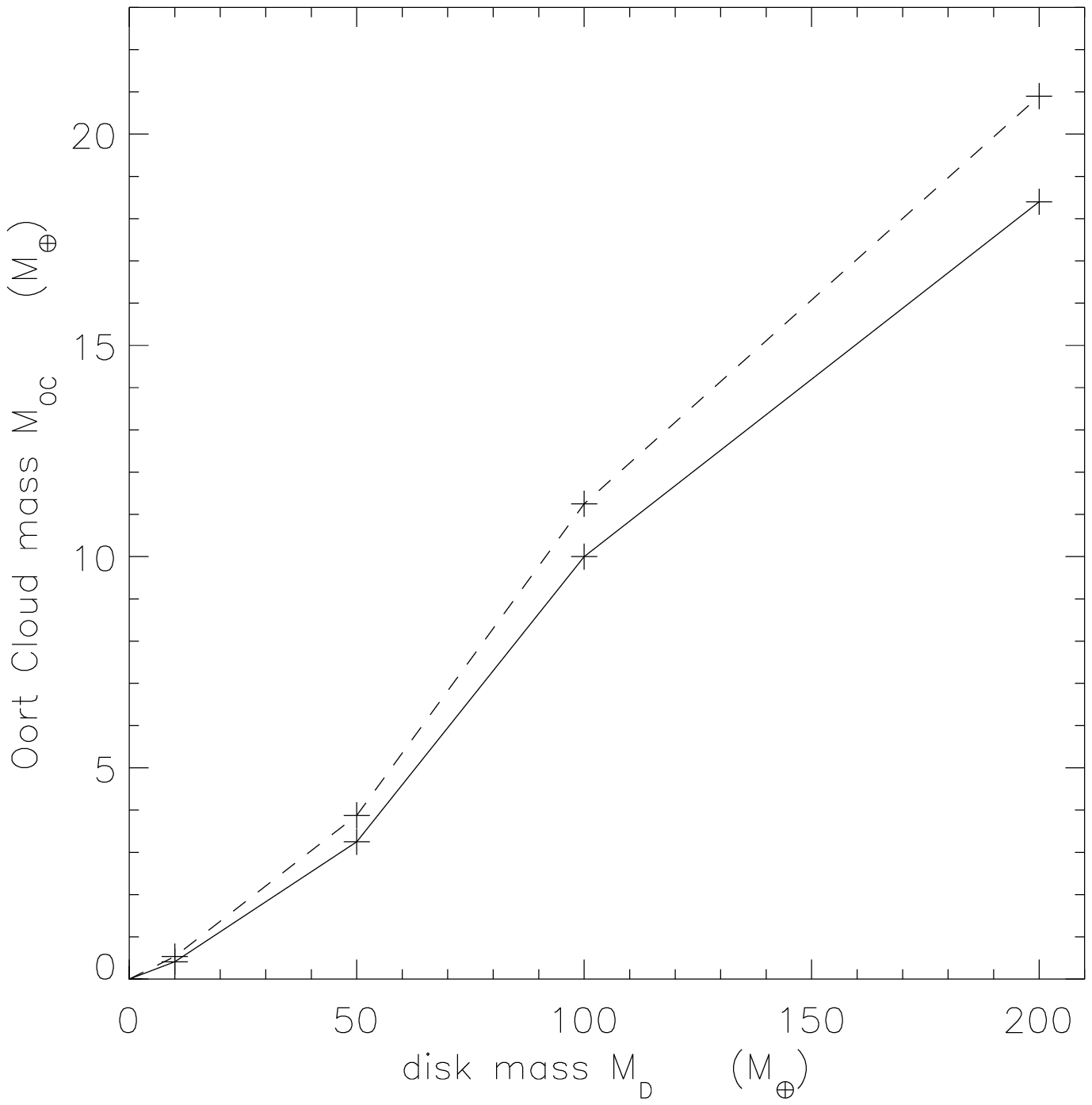,height=6.0in,width=6.0in}
Figure \ref{Moc}
\end{figure}

\clearpage
\begin{figure}[t]
\psfig{figure=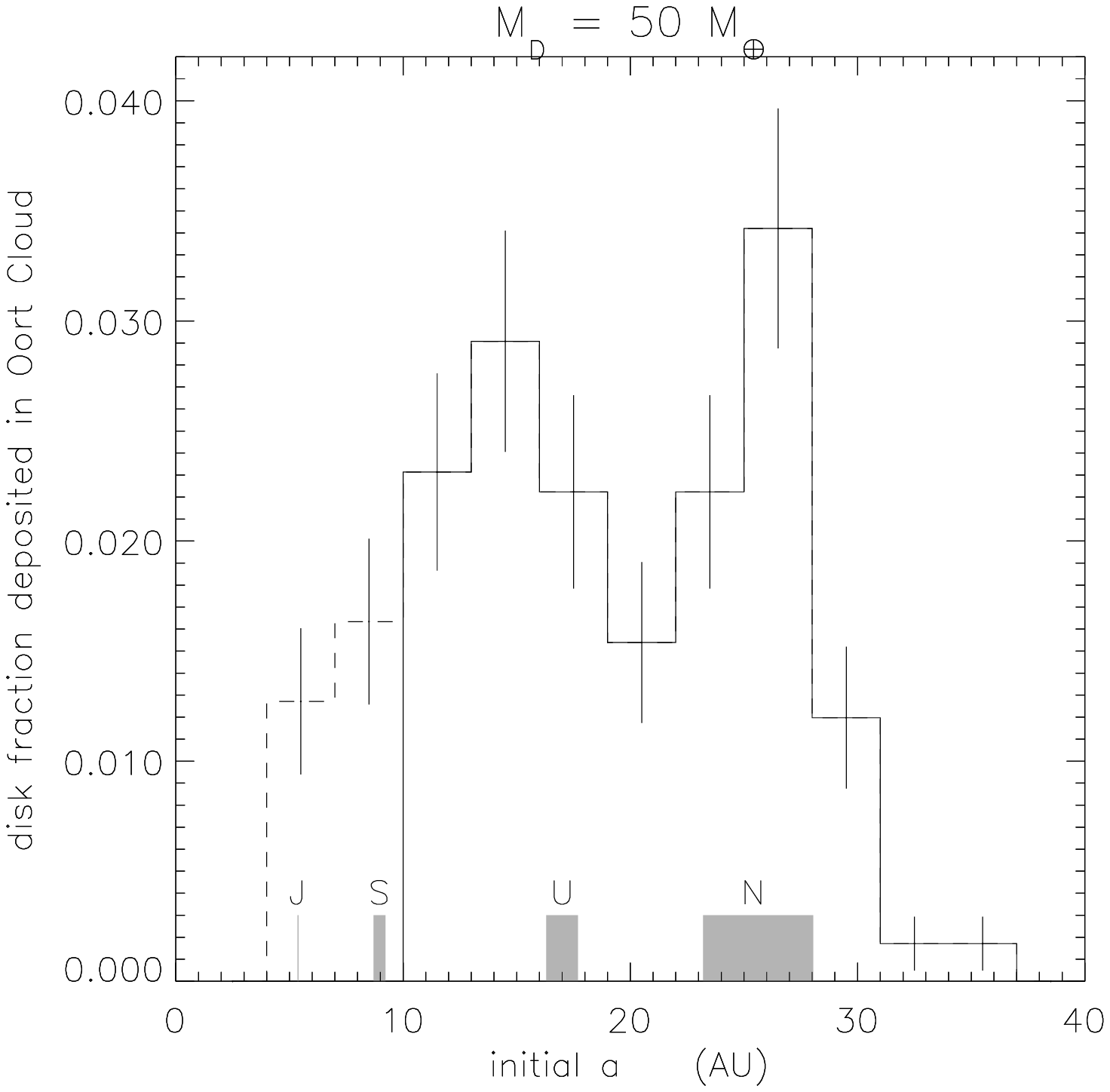,height=6.0in,width=6.0in}
Figure \ref{OC_origin}
\end{figure}

\clearpage
\begin{figure}[t]
\psfig{figure=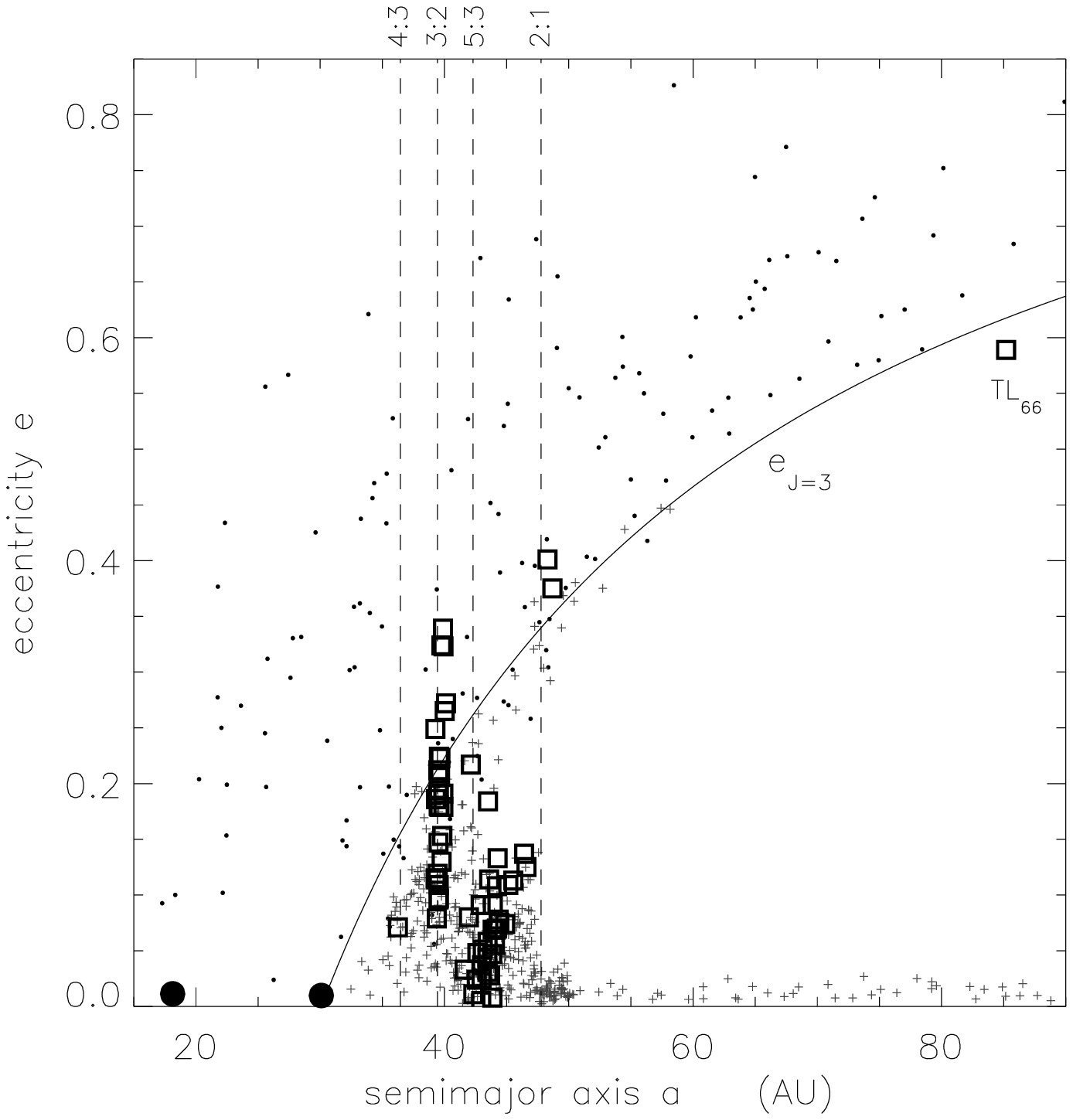,height=6.0in,width=6.0in}
Figure \ref{disk}
\end{figure}


\newpage
\begin{thebibliography}{99}

\bibitem[Dermott, Malhotra \& Murray 1988]{DMM88}
  Dermott, S.\ F., Malhotra, R., \& Murray, C.\ D. 1988, Icarus,
  76, 295

\bibitem[Duncan, Quinn, \& Tremaine 1987]{DQT87}
  Duncan, M., Quinn, T., \& Tremaine, S. 1987, AJ, 94, 1330

\bibitem[Duncan, Levison, \& Budd 1995]{DLB95}
  Duncan, M.\ J., Levison, H.\ F., \& Budd, S.\ M. 1995,
  AJ, 110, 3073

\bibitem[Duncan \& Levison 1997]{DL97}
  Duncan, M.\ J.\ \& Levison, H.\ F. 1997, Science, 276, 1670

\bibitem[Duncan, Levison, \& Lee 1998]{DLL98}
  Duncan, M.\ J., Levison, H.\ F., \& Lee, M.\ H. 1998,
  AJ, 116, 2067

\bibitem[Everhart 1967]{E67}
  Everhart, E. 1967, AJ, 72, 1002

\bibitem[Fern\'{a}ndez \& Ip 1984]{FI84}
  Fern\'{a}ndez, J.\ A.\ \& Ip, W.--H. 1984, Icarus, 58, 109

\bibitem[Fern\'{a}ndez \& Ip 1996]{FI96}
  Fern\'{a}ndez, J.\ A.\ \& Ip, W.--H. 1996,
  Planet.\ Space Sci., 44, 431

\bibitem[Goldreich \& Tremaine 1980]{GT80}
  Goldreich, P.\ \& Tremaine, S. 1980, Ap.\ J., 241, 425

\bibitem[Greaves {\it et al.} 1998]{Greaves98}
  Greaves, J.\ S., Holland, W.\ S., Moriarty--Schieven, G.,
  Jeness, T., Dent, W.\ R.\ F., Zuckerman, B., McCarthy, C.,
  Webb, R.\ A., Butner, H.\ M., Gear, W.\ K., \& Walker, H.\ J.
  1998, Ap.\ J., 506, L133

\bibitem[Hahn, Ward, \& Rettig 1995]{HWR95}
  Hahn, J.\ M., Ward, W.\ R., \& Rettig, T.\ W.
  1995, Icarus, 117, 25

\bibitem[Heisler 1990]{H90}
  Heisler, J. 1990, Icarus, 88, 104

\bibitem[Holland {\it et al.} 1998]{Hetal98}
  Holland, W.\ S., Greaves, J.\ S., Zuckerman, B., Webb, R.\ A.,
  McCarthy, C., Coulson, I.\ M., Walther, D.\ M., Dent, W.\ R.\ F.,
  Gear, W.\ K., \& Robson, I. 1998, Nature, 392, 788
 
\bibitem[Ida \& Makino 1993]{IM93}
  Ida, S.\ \& Makino, J. 1993, Icarus, 106, 210

\bibitem[Jewitt, Luu, \& Trujillo 1998]{Jetal98}
  Jewitt, D., Luu, J., \& Trujillo, C. 1998, AJ, 115, 2125

\bibitem[Kenyon \& Luu 1998]{KL98}
  Kenyon, S.\ \& Luu, J.\ X. 1998, AJ, 115, 2136

\bibitem[Koerner {\it et al.} 1998]{Ketal98}
  Koerner, D.\ W., Ressler, M.\ E., Werner, M.\ W., \& Backman, D.\ E.
  1998, Ap.\ J., 503, L83

\bibitem[Kozai 1962]{K62}
  Kozai, Y. 1962, AJ, 67, 591

\bibitem[Levison 1996]{L96}
  Levison, H.\ F. 1996. In ASP Conf.\ Ser. 107, Completing the
  Inventory of the Solar System, eds. T.\ W.\
  Rettig \& J.\ M.\ Hahn (San Francisco: ASP), 173

\bibitem[Levison \& Duncan 1994]{LD94}
  Levison, H.\ F.\ \& Duncan, M.\ J. 1994, Icarus, 108, 18

\bibitem[Levison \& Stern 1995]{LS95}
  Levison, H.\ F.\ \& Stern, S.\ A. 1995, Icarus, 116, 315

\bibitem[Lissauer \& Espresate 1998]{LE98}
  Lissauer, J.\ J. \& Espresate, J. 1998, Icarus, 134, 155

\bibitem[Lynden--Bell \& Kalnajs 1972]{LK72}
  Lynden--Bell, D. \& Kalnajs, A.\ J. 1972, Mon.\ Not.\ R.\ Astr.\
  Soc., 157, 1

\bibitem[Lynden--Bell \& Pringle 1974]{LP74}
  Lynden--Bell, D.\ \& Pringle, J.\ E. 1974, Mon.\ Not.\ R.\ Astr.\
  Soc., 168, 603

\bibitem[Malhotra 1993]{RM93}
  Malhotra, R. 1993, Nature, 365, 819

\bibitem[Malhotra 1995]{RM95} 
  Malhotra, R. 1995, AJ, 110, 420

\bibitem[Malhotra (1997)]{RM97}
  Malhotra, R. 1997, Planetary and Space Science, in press

\bibitem[Malhotra 1998]{RM98}
  Malhotra, R. 1998, Lunar and Planetary Science XXIX, 1476

\bibitem[Marsden 1998]{BM98} Marsden 1998. From the IAU Minor Planet
  Center,\\ http://cfa-www.harvard.edu/iau/lists/TNOs.html

\bibitem[Morbidelli \& Valsecchi 1997]{MV97}
  Morbidelli, A.\ \& Valsecchi, G.\ B. 1997, Icarus, 128, 464

\bibitem[\"{O}pik (1951)]{O51}
  \"{O}pik, E.\ J. 1951, Proc.\ R.\ I.\ A., 54, 25

\bibitem[Rauch \& Holman 1999]{RH98}
  Rauch, K.\ P.\ \& Holman, M. 1999, AJ, in press

\bibitem[Saha \& Tremaine 1992]{ST92}
  Saha, P.\ \& Tremaine, S. 1992, AJ, 104, 1633

\bibitem[Shoemaker \& Wolfe 1984]{SW84}
  Shoemaker, E.\ M.\ \& Wolfe, R.\ F. 1984, Lunar and Planetary
  Science XV, 780

\bibitem[Smith \& Terrile 1984]{ST84}
  Smith, B.\ A.\ \& Terrile, R.\ J. 1984, Science, 226, 1421

\bibitem[Smith {\it et al.} 1998]{Setal98}
  Smith, B.\ A., Schneider, G., Becklin, E.\ E., Koerner, D.\ W.,
  Meier, R., Terrile, R.\ J., Hines, D.\ C., Lowrance, P.\ J., \&
  Thompson, R.\ I. 1998, BAAS, 30, 1382

\bibitem[Stern \& Colwell 1997]{SC97}
  Stern, S.\ A.\ \& Colwell, J.\ E. 1997, Ap.\ J., 490, 879

\bibitem[Toomre 1969]{T69}
  Toomre, A. 1969, Ap.\ J., 158, 899

\bibitem[Trilling \& Brown 1998]{TB98}
  Trilling, D.\ E.\ \& Brown, R. 1998, Nature, 395, 775

\bibitem[Ward 1981]{W81}
  Ward, W.\ R. 1981, Icarus, 47, 234

\bibitem[Weissman 1996]{W96}
  Weissman, P.\ R. 1996. In ASP Conf.\ Ser. 107, Completing the
  Inventory of the Solar System, eds. T.\ W.\
  Rettig \& J.\ M.\ Hahn (San Francisco: ASP), 265

\bibitem[Weissman \& Levison (1997)]{WL97}
  Weissman, P.\ R.\ \& Levison, H.\ F. 1997, Ap.\ J., 488, L133

\bibitem[Wetherill 1975]{W75}
  Wetherill, G.\ W. 1975, Proc.\ $6th$ Lunar Sci.\ Conf., 1539

\bibitem[Wisdom \& Holman 1991]{WH91}
  Wisdom, J.\ \& Holman, M. 1991, AJ, 102, 1528


\bibitem[Duncan, Quinn, \& Tremaine (1987)]{DQT(87)}
\bibitem[Duncan \& Levison (1997)]{DL(97)}
\bibitem[Duncan, Levison, \& Lee (1998)]{DLL(98)}
\bibitem[Fern\'{a}ndez \& Ip (1984)]{FI(84)}
\bibitem[Levison \& Duncan (1994)]{LD(94)}
\bibitem[Morbidelli \& Valsecchi (1997)]{MV(97)}
\bibitem[Saha \& Tremaine (1992)]{ST(92)}
\bibitem[Weissman (1996)]{W(96)}
\bibitem[Wisdom \& Holman (1991)]{WH(91)}

\end{thebibliography}
\end{document}